\documentclass[floats,floatfix,amssymb,prd,twocolumn,superscriptaddress,nofootinbib,preprintnumbers]{revtex4-1}

\usepackage{subcaption}
\usepackage{ragged2e}
\DeclareCaptionJustification{justified}{\justifying}
\makeatletter
\newcommand{\subsetsim}{\mathrel{\mathpalette\subset@sim\relax}}
\newcommand{\subset@sim}[2]{%
  \vtop{\offinterlineskip\m@th
    \ialign{\hfil##\cr
      $#1\subset$\cr\noalign{\kern0.5pt}\scalebox{0.9}{$#1\sim$}\cr
    }%
  }%
}
\makeatother

\usepackage{amssymb,amsmath,verbatim,mathtools,needspace,enumitem,etoolbox,graphicx,microtype,afterpage,bm}

\usepackage[dvipsnames, usenames]{xcolor}

\definecolor{linkcolor}{rgb}{0.0,0.3,0.5}
\usepackage{booktabs}
\usepackage[unicode, colorlinks=true, linkcolor=linkcolor, citecolor=linkcolor, filecolor=linkcolor,urlcolor=linkcolor, pdfusetitle]{hyperref}
\usepackage[all]{hypcap}
\usepackage[T1]{fontenc}
\usepackage[utf8]{inputenc}
\usepackage{tabularx}
\usepackage{float}
\renewcommand{\arraystretch}{1.4}

\usepackage{multirow}
\usepackage{pifont}
\usepackage{lmodern}

\allowdisplaybreaks
\usepackage{tikz}
\usetikzlibrary{shapes.misc, positioning, arrows.meta}
\usepackage{framed}
\usepackage{hyperref}
\hypersetup{colorlinks, citecolor=bluscuro, linkcolor=black, urlcolor=bluscuro}
\definecolor{rossos}{cmyk}{0,1,1,0.55}
\definecolor{bluscuro}{rgb}{0.15, 0.2, .85}
\definecolor{bluchiaro}{cmyk}{1,.3,0.,0.1}
\definecolor{ForestGreen}{rgb}{0.13, 0.55, 0.13}
\definecolor{TLGreen}{RGB}{50, 164, 49}
\definecolor{TLOrange}{RGB}{231,180,22}
\definecolor{TLRed}{RGB}{204,50,50}

\newcommand{\TLBullet}[1]{\raisebox{-5pt}{\scalebox{0.23}{\begin{tikzpicture}\shadedraw[rounded corners=15pt, top color=gray!84!black,bottom color=black, line width=.6pt] (0,0) rectangle ++(6,2); \ifthenelse{#1=1}{\draw[fill=green,line width=1.pt]  (1,1) circle(.75cm);}{\draw[fill=green!35!black,line width=1.pt]  (1,1) circle(.75cm);}\ifthenelse{#1=2}{\draw[fill=yellow,line width=1.pt]  (3,1) circle(.75cm);}{\draw[fill=yellow!60!black,line width=1.pt]  (3,1) circle(.75cm);}\ifthenelse{#1=3}{\draw[fill=red,line width=1.pt]  (5,1) circle(.75cm);}{\draw[fill=red!50!black,line width=1.pt]  (5,1) circle(.75cm);}\end{tikzpicture}}}}
\newcommand{\apjl}{"Astrophys. J. Lett."}

\def\d{{\mathrm{d}}}

\newcommand{\bs}{\begin{subequations}}
\newcommand{\es}{\end{subequations}}

\newcommand{\be}{\begin{equation}}
\newcommand{\ee}{\end{equation}}
\renewcommand{\d}{{\rm d}}

\newcommand{\lp}{\left (}
\newcommand{\rp}{\right )}

\def\lsim{\mathrel{\rlap{\lower4pt\hbox{\hskip0.5pt$\sim$}}
    \raise1pt\hbox{$<$}}}         
\def\gsim{\mathrel{\rlap{\lower4pt\hbox{\hskip0.5pt$\sim$}}
    \raise1pt\hbox{$>$}}}         

\usepackage{siunitx}
\DeclareSIUnit \parsec {pc}
\DeclareSIUnit \arcsecondfull {arcsec}
\DeclareSIUnit \year{yr}
\DeclareSIUnit \day{day}
\DeclareSIUnit \hour{hr}
\DeclareSIUnit \radiant{rad}
\DeclareSIUnit \degfull{deg}
\DeclareSIUnit \erg {erg}
\DeclareSIUnit \Lsun {L_\odot}
\DeclareSIUnit \Msun {M_\odot}
\DeclareSIUnit \AstroUnit {au}

\usepackage{nicefrac}

\newcommand{\sapienza}{Dipartimento di Fisica, Sapienza Università 
	di Roma, Piazzale Aldo Moro 5, 00185, Roma, Italy}
\newcommand{\infn}{INFN, Sezione di Roma, Piazzale Aldo Moro 2, 00185, Roma, Italy}

\newcommand{\unige}{D\'epartement de Physique Th\'eorique,
Universit\'e de Gen\`eve, 24 quai Ansermet, CH-1211 Gen\`eve 4, Switzerland}
\newcommand{\gwsc}{Gravitational Wave Science Center (GWSC), Universit\'e de Gen\`eve, CH-1211 Geneva, Switzerland}

\newcommand{\cern}{
CERN, Theoretical Physics Department,
Esplanade des Particules 1, Geneva 1211, Switzerland}

\begin{document}

\title{Can we identify primordial black holes?\\ Tidal tests for subsolar-mass gravitational-wave observations}

\author{Francesco Crescimbeni}
\email{francesco.crescimbeni@uniroma1.it}
\affiliation{\sapienza}
\affiliation{\infn}

\author{Gabriele Franciolini}
\email{gabriele.franciolini@cern.ch}
\affiliation{\cern} 

\author{Paolo Pani}
\email{paolo.pani@uniroma1.it}
\affiliation{\sapienza}
\affiliation{\infn}

\author{Antonio Riotto}
\email{antonio.riotto@unige.ch}
\affiliation{\unige}
\affiliation{\gwsc}

\date{\today}

\begin{abstract}
\noindent
The detection of a subsolar object in a compact binary merger is regarded as one of the smoking gun signatures of a population of primordial black holes~(PBHs). We critically assess whether these systems could be distinguished from stellar binaries, for example composed of white dwarfs or neutron stars, which could also populate the subsolar mass range. At variance with PBHs, the gravitational-wave signal from stellar binaries is affected by tidal effects, which dramatically grow for moderately compact stars as those expected in the subsolar range. We forecast the capability of constraining tidal effects of putative subsolar neutron star binaries with current and future LIGO-Virgo-KAGRA (LVK) sensitivities as well as next-generation experiments. We show that, should LVK O4 run observe subsolar neutron-star mergers, it could measure the (large) tidal effects with high significance. In particular, for subsolar neutron-star binaries, O4 and O5 projected sensitivities would allow measuring the effect of tidal disruption on the waveform in a large portion of the parameter space, also constraining the tidal deformability at ${\cal O}(10\%)$ level, thus excluding a primordial origin of the binary. Vice versa, for subsolar PBH binaries, model-agnostic upper bounds on the tidal deformability can rule out neutron stars or more exotic competitors. Assuming events similar to the subthreshold candidate SSM200308 reported in LVK O3b data are PBH binaries, O4 projected sensitivity would allow ruling out the presence of neutron-star tidal effects at $\approx 3 \sigma$ C.L., thus strengthening the PBH hypothesis. Future experiments would lead to even stronger ($>5\sigma$) conclusions on potential discoveries of this kind.

\end{abstract}

\preprint{ET-0060A-24. CERN-TH-2024-026.}

\maketitle

\section{Introduction}\label{intro}

The observation of a subsolar mass (SSM) object in a binary black hole~(BBH) merger is considered as the most robust smoking gun of the primordial nature of a binary \cite{Green:2020jor}. 
Exploiting this window, however, relies on our capabilities to distinguish such event from other astrophysical systems and other potential candidates from new physics~\cite{Franciolini:2021xbq,Cardoso:2019rvt}.

A signal compatible with a subsolar merger could be observed already during the ongoing O4 run of the LIGO/Virgo/Kagra (LVK) Collaboration. 
Previous LVK observation campaigns reported the existence of SSM candidate events with too low significance to be classified as confident detections~\cite{LIGOScientific:2022hai}, and were not included in the LVK merger catalog~\cite{KAGRA:2021duu}.
Subsequent work, especially Ref.~\cite{Prunier:2023cyv}, reported the analysis of the SSM candidate
SSM200308, possibly composed by two subsolar BHs with masses 
$m_1= 0.62^{+0.46}_{-0.20}M_\odot$ and
$m_2= 0.27^{+0.12}_{-0.10} M_\odot$ at a redshift of $z=0.02^{+0.01}_{-0.01}$ (90\% C.I.), showing relatively small errors on the determination of both masses even for such subthreshold event.
This is because light mergers perform a large number of cycles in the detector band. 
However, the SSM nature of the event alone is not sufficient to claim the observation of a PBH binary.
Since a robust detection of a subsolar BH\footnote{Population studies suggest that, if some of the O3 events are PBHs, there will be a non-negligible probability to detect subsolar events starting from O4~\cite{Franciolini:2022tfm}.} would be a breakthrough with a strong impact on cosmology, high-energy physics, and astrophysics~\cite{LISACosmologyWorkingGroup:2023njw,Carr:2020xqk,Sasaki:2018dmp}, it is of utmost importance to exclude any possible source of confusion that might affect such a putative detection.

In this paper, we investigate whether current and future experiments would have sufficient sensitivity required to detect SSM200308-like events, and distinguish this from other astrophysical systems or more exotic competitors in the SSM range.

Although standard formation scenarios suggest that astrophysical compact objects have typically masses above $M_\odot$ [and below a critical mass $\sim {\cal O}(M_\odot)$], both white dwarfs~(WDs) and neutron stars~(NSs) can in principle be subsolar. In typical astrophysical settings, WDs are formed with masses as low as $\approx 0.2M_\odot$~\cite{Kilic:2006as}.
NSs are observed through x-ray observations and GWs~\cite{Ozel:2016oaf,Riley:2019yda,LIGOScientific:2018cki,De:2018uhw,Most:2018hfd}. All these observations point toward a population of supersolar mass NSs, although there might be selection biases, especially for x-ray sources. Smaller masses are possible for cold, dense equations of state~(EoS)~\cite{Lattimer:2000nx}, even though very light NSs are unstable to expansion~\cite{Colpi1991} and
supernova theory suggests heavier lower bounds on their mass exits~\cite{1983bhwd.book.....S,Suwa:2018uni}. 
Examples of such low mass objects may have already been observed, see e.g. HESS~J1731–347~\cite{2022NatAs...6.1444D} (a candidate NS with mass $0.77^{+0.20}_{-0.17} M_\odot$) and the candidate object reported in Ref.~\cite{2019ApJ...881L...3M} (possibly a white dwarf with mass $0.20^{+0.01}_{-0.01} M_\odot$).

Furthermore, material compact objects other than WDs and NSs can exist in the landscape of beyond-Standard-Model physics~\cite{Giudice:2016zpa,Cardoso:2019rvt}, and might possibly have a cosmological origin. Notable examples include Q-balls~\cite{Coleman:1985ki}, boson stars~\cite{Liebling:2012fv}, fermion-soliton stars~\cite{Lee:1986tr,DelGrosso:2023trq,DelGrosso:2023dmv} (see~\cite{Cardoso:2019rvt} for an overview); most these models can accommodate subsolar compact objects, depending on the coupling of the underlying fundamental theory~\cite{DelGrosso:2024wmy}.

\begin{figure}[t!]
\centering
\includegraphics[width=0.49\textwidth]{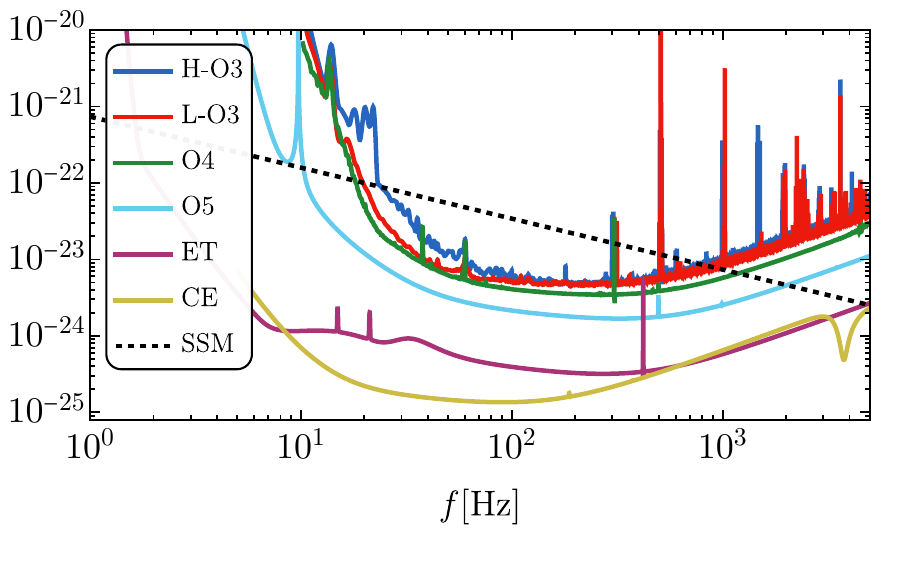}
\caption{ 
Current and future sensitivity curves for Livingston and Handford LIGO experiments during O3, O4, and O5 observations runs, alongside next-generation Einstein Telescope (ET) and Cosmic Explorer (CE). 
The black dashed line indicates the GW amplitude ($2 |\tilde h (f)| \sqrt{f}$) for the inspiral phase of a SSM200308-like merger. The ISCO frequency sits outside the range shown here.
}\label{fig:sens}
\end{figure}

A key difference between any material compact object and a BH is that only in the latter case does the tidal deformability (as measured by the so-called tidal Love numbers) vanishes~\cite{Damour_tidal, Binnington:2009bb, Damour:2009vw,Chia:2020yla}. This is a unique property of BHs in general relativity which can be understood in terms of special symmetries (see, e.g.,~\cite{Charalambous:2021kcz}). However, this property does not hold for any other material object~\cite{Porto:2016zng, Cardoso:2017cfl}. 

Although the tidal Love numbers affect the GW signal only at high post-Newtonian~(PN) order~\cite{Flanagan:2007ix,Hinderer:2009ca}, they strongly depend on the compactness and grows dramatically for less compact objects as those expected, for example, in the subsolar tail of a NS mass-radius diagram~\cite{Ozel:2016oaf}.
In this case, the tidal deformability can be orders of magnitude larger than for the ordinary compact NSs detected so far, magnifying tidal effects in the waveform and making them potentially measurable even for systems that are detected mostly in the early inspiral (see Fig.~\ref{fig:sens}).
Furthermore, less compact objects may be disrupted during the inspiral. This would cause the GW signal to be damped much before the maximum frequency expected for objects as compact as BHs. This feature may also be used to test the nature of the observed binary.

The main scope of this paper is to assess whether the tidal deformability and the disruption frequency can be used to confirm/rule out the PBH origin of a SSM GW event.

\section{Tidal deformability and disruption tests of SSM signals}

\subsection{Binary maximum frequency of material compact objects}
The waveform of a signal emitted by a binary BH is characterized by a maximum signal frequency which is of the order of the  innermost stable circular orbit (ISCO) frequency, defined as 
\begin{equation}\label{eq:fisco}
f_\text{\tiny ISCO} 
= \frac{c^3}{(6^{3/2}\pi G M)}
= 4.4\,  {\rm kHz} \left ( M_\odot \over {M}\right ).
\end{equation}
with $M=m_1 + m_2$ being the total mass of the binary.

However, binaries of stellar objects are typically characterized by smaller maximal frequencies, either because they have a hard surface and their contact frequency is smaller than $f_\text{\tiny ISCO}$ or because the least compact companion can be tidally disrupted during the inspiral. We can provide a rough estimate of the tidal disruption radius $r_{T}$ by equating the tidal force and the object self-gravity~\cite{1983bhwd.book.....S},
\begin{equation}
    r_{T,i} = \left( \frac{2 m_j}{m_i} \right)^{1/3} r_{i}\,,
\end{equation}
where $i,j=1,2$ are indices denoting the two objects, and $r_i$ is the radius of the $i$th object.
When the binary reaches the largest $r_{T,i}$, the lighter object can be considered tidally disrupted. 
This corresponds to a frequency of 
\begin{equation} \label{fT}
    f_T 
    = 
    \frac{1}{\pi} \sqrt{\frac{G M}{
    (  \max[r_{T,1}, r_{T,2}]
    )^3}}\,.
\end{equation}
Let us first discuss the case of WDs and then turn to the more interesting case of NSs. 

\subsubsection{White dwarfs}
\indent
Potential WD-WD binaries could be distinguished from more compact subsolar binaries based on their lower maximum frequency. 
Assuming a WD mass-radius relation of the form~\cite{Magano:2017mqk}
\begin{equation}
    r_\text{\tiny WD} = 0.013 \,
    r_\odot \left(\frac{m_\text{\tiny WD}}{M_\odot}\right)^{-1/3}\,,
\end{equation}
and nearly equal mass binaries, we find that 
\begin{equation}
    f_\text{\tiny max}^{\text{\tiny WD}} = 0.13 \,{\rm Hz} \left ( \frac{m_\text{\tiny WD}}{ M_\odot}\right ).
\end{equation}
\noindent
If the primary object is heavier than a WD (for example, an ordinary, supersolar, BH), the above frequency is smaller. In the $m_1\gg m_\text{\tiny WD}$ limit, $f_\text{\tiny max}^{\text{\tiny WD}}\sim 0.095 (m_\text{\tiny WD}/M_\odot)\,{\rm Hz}$.
Therefore, we do not expect WD binaries to contaminate the range of frequencies observable by ground-based detectors ($f\gtrsim {\cal O}({\rm few)}\, $Hz).
Detection prospects to distinguish WD binaries from 
possible mixed WD-PBH binaries using deci-Hz GW detectors were recently discussed in~\cite{Yamamoto:2023tsr}.

\subsubsection{Neutron stars}\label{sec:maxfNS}
Assuming an equal-mass NS binary, from Eq.~\eqref{fT} we obtain
\begin{equation}
    f_\text{\tiny max}^{\text{\tiny NS}} \approx 1.4 \,{\rm kHz} \left ( \frac{m_\text{\tiny NS}}{ 0.5M_\odot}\right )^{1/2} \left (\frac{15\,{\rm km}}{ r_\text{\tiny NS}}\right )^{3/2}.
\end{equation}
where we normalized the NS mass and radius to typical subsolar values. Also in this case, if $m_1\gg m_\text{\tiny NS}$, the above frequency is slightly smaller, $f_\text{\tiny max}^{\text{\tiny NS}} \approx 1 \,{\rm kHz} \left ( \frac{m_\text{\tiny NS}}{ 0.5M_\odot}\right )^{1/2} \left (\frac{15\,{\rm km}}{ r_\text{\tiny NS}}\right )^{3/2}.$
In this case, the maximum frequency is within or above the bandwidth of ground-based detectors, so the inspiral of these binaries is potentially detectable.

One can also adopt more accurate estimates for the maximum merger frequencies of NSs. 
As found in numerical simulations, the binary NS merger is expected to occur shortly after the Roche overflow of the secondary star. 
Therefore, Roche lobe overflow can be used as a conservative time at which to terminate an inspiral gravitational waveform model \cite{Shibata:2001ag, Marronetti:2003hx,Dietrich:2015pxa, Bernuzzi:2020txg}.
Reference~\cite{Bandopadhyay:2022tbi} performed numerical simulations to derive $f_\text{\tiny RO}$, which can be analytically approximated as\footnote{We use the data contained in the repository \href{https://github.com/sugwg/sub-solar-ns-detectability}{https://github.com/sugwg/sub-solar-ns-detectability} \cite{Bandopadhyay:2022tbi}.}
\begin{align}\label{eq:RO}
    f_\text{\tiny RO} / {\rm Hz}
    =
    -26.9 - 35.5\left ( \frac{m_1}{M_\odot} \right)
    - 3.02\left ( \frac{m_1}{M_\odot}\right)^2 
    \nonumber 
    \\
    + 1690 \left (\frac{m_2}{M_\odot} \right ) 
    - 575 \left ( \frac{m_2}{M_\odot} \right )^2.
\end{align}
as a function of the binary masses. This fit reproduces the numerical results with errors $<10\%$ in the range of masses $[0.2,1] M_\odot$ and assumes APR EoS, even though it is not very sensitive to this choice, at least for the cases tested in Ref.~\cite{Bandopadhyay:2022tbi}.

While neglecting disruption would impact the search sensitivity~\cite{Cullen:2017oaz,Bandopadhyay:2022tbi}, for a given detected signal having disruption before the ISCO would exacerbate the difference between the point-particle inspiral waveform and the actual signal, thus providing additional information. 
To include this effect, one can introduce a frequency-dependent smoothing of the GW signal following the disruption of binary components. 
We model this effect by adopting a phenomenological tapering function (see ~\cite{Maselli:2013rza} for a similar model)\footnote{See also~\cite{DeLuca:2022xlz} for a similar phenomenological model recently introduced to incorporate frequency-dependent tidal disruption motivated by environmental effects.}
\begin{equation}
  {\cal T}(f|f_\text{\tiny cut},f_\text{\tiny slope}) =  
\left [
\frac{1+e^{-f_\text{\tiny cut}/f_\text{\tiny slope}}}{1+e^{(f-f_\text{\tiny cut})/f_\text{\tiny slope}}}
\right ],
\end{equation}
where 
$f_\text{\tiny cut}$ and $f_\text{\tiny slope}$ parametrize the location and rapidity of the suppression of the signal, respectively. In practice, we will account for the effect of tidal disruption in the waveform by convoluting the tapering function with an ordinary waveform:
\begin{equation}
    \tilde h (f) \to \tilde h (f) \, {\cal T}(f|f_\text{\tiny cut},f_\text{\tiny slope})\,,
\end{equation}
where $\tilde h$ is the Fourier transform of the signal.

\subsection{Tidal deformability}
When at least one of the binary components is a material object, the GW phase emitted during the inspiral depends on the effective tidal deformability parameter~\cite{Flanagan:2007ix,Hinderer:2009ca}
\begin{equation} \label{Lambda}
    \Lambda = 
    \frac{2}{3} k_2
    \left( \frac{G m}{R} \right)^{-5},
\end{equation}
where $k_2$ is the quadrupolar tidal Love number (which identically vanishes if the object is a BH), $R$ is the object radius, and $m$ is its mass.

\subsubsection{Neutron stars}
To account for the uncertainties in the NS EoS at ultranuclear densities, we assume three different EoS models, namely the APR~\cite{Akmal:1998cf}, SLy4~\cite{Douchin:2001sv}, and BSk21~\cite{Pearson:2011zz,Pearson:2012hz,Potekhin:2013qqa} EoS. These models all use a unified nonrelativistic formalism and produce NSs that are consistent with the maximum mass of known pulsars~\cite{NANOGrav:2019jur,Barr:2024wwl}, the measured values of NS radii from NICER, and the values of $\Lambda$ which are compatible with those measured in GW170817~\cite{LIGOScientific:2018cki,De:2018uhw,Most:2018hfd,Radice:2018ozg,Raithel:2018ncd,Capano:2019eae,Riley:2019yda, Miller:2019cac,Miller:2021qha,Al-Mamun:2020vzu}. 

\begin{figure}[t!]
\centering
\includegraphics[width=0.49\textwidth]{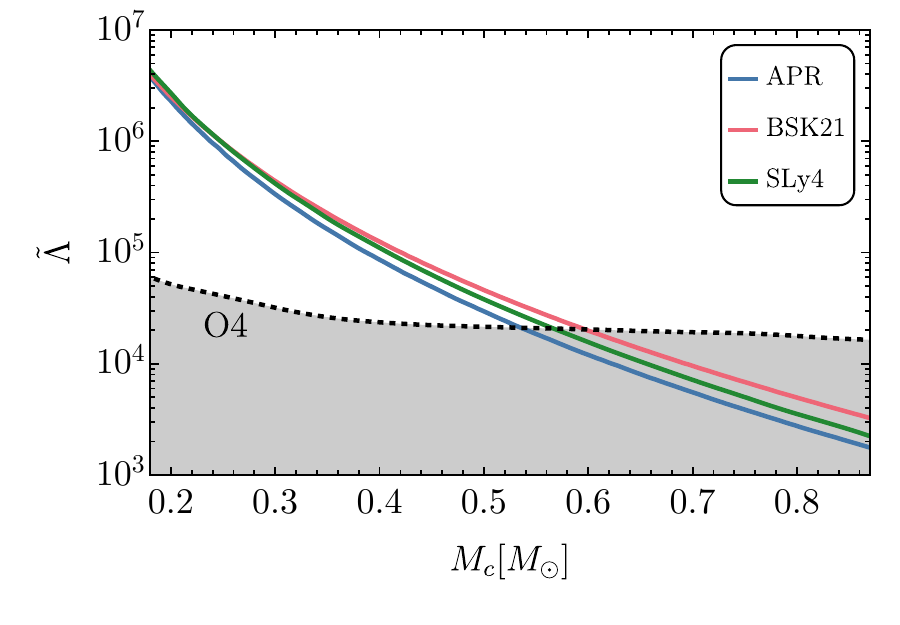}
\caption{ 
Effective deformability parameter for a NS binary with three different EoS. 
We show results varying the chirp mass $M_c$ assuming equal mass binaries.
The dashed black line indicates the upper bound (at $3 \sigma$) obtained for an equal mass binary with SNR = 12 at LVK O4.
}\label{fig:tL}
\end{figure}

In Fig.~\ref{fig:tL} we show the effective tidal deformability parameter of a binary (defined in the next section) as a function of the chirp mass for equal-mass binaries and different EoS. As expected, the magnitude of this parameter is not very sensitive to the EoS, especially in the low-mass regime, where stars are less and less compact and deviations among different EoS are smaller.\footnote{
One can estimate an analytical scaling of the tidal deformability in the Newtonian regime. For a polytropic fluid with pressure-density relation $P\propto \rho^{1+1/n}$, the mass-radius relation reads $m\propto R^{\frac{3-n}{1-n}}$ and the tidal Love number is constant $k_2={\cal O}(1)$~\cite{Binnington:2009bb} Using Eq.~\eqref{Lambda}, one finds $\Lambda\propto m^{-10/(3-n)}$, yielding $\Lambda\propto m^{-6.66}$ for a nonrelativistic degenerate gas ($n=3/2$)~\cite{1983bhwd.book.....S}, whereas $\Lambda\propto m^{-5}$ for $n=1$.
}

For individual components, assuming for example SLy4 EoS, one finds that numerical data are well fitted by
\begin{equation}
    \Lambda = 7.3 \cdot 10^4 \left ( m \over 0.5 M_\odot \right)^{-4.7}. \label{Sly4}
\end{equation}

\subsubsection{Exotic compact objects}
In various scenarios beyond the Standard Model, exotic compact objects other than BHs or NSs can exist and might also have a primordial origin~\cite{Giudice:2016zpa,Cardoso:2019rvt}. These models can be more compact and more massive than ordinary NSs and are intensively studied as regular BH mimickers~\cite{Cardoso:2019rvt,Maggio:2021ans}. The tidal Love numbers of an exotic compact object were explicitly computed for different models such as boson stars~\cite{Cardoso:2017cfl,Sennett:2017etc,Mendes:2016vdr}, fermion-boson stars~\cite{Diedrichs:2023trk}, gravastars~\cite{Pani:2015tga,Cardoso:2017cfl,Uchikata:2016qku}, anisotropic stars~\cite{Raposo:2018rjn} and other simple models with stiff EoS at the surface~\cite{Cardoso:2017cfl}. As expected, the TLNs are generically nonzero and vanish in the BH limit~\cite{Pani:2015tga,Cardoso:2017cfl} (for those models in which such limit exists).

Arguably the best motivated and studied model of exotic compact objects are boson stars, which are solutions to the Einstein-Klein-Gordon system with scalar self-interactions~\cite{Liebling:2012fv}. The tidal Love numbers of a boson star depend strongly on its compactness and on the scalar potential of the underlying theory, and can range within several orders of magnitude~\cite{Cardoso:2017cfl,Sennett:2017etc,Mendes:2016vdr,Pacilio:2020jza}. As a general rule of thumb, the stronger the scalar self-interaction the higher the maximum compactness and the smallest the tidal Love number of the maximum mass configuration. The minimal model of boson star (with a mass term and no self interactions) has a Love number that is bigger than in the typical NS case, whereas more interacting scalar field theories can support solutions with smaller Love numbers.

For example, for boson stars with a quartic potential $V\left(\left|\phi\right|\right)=\frac{\mu^2}{2}\left|\phi\right|^2+\frac{\lambda}{4}\left|\phi\right|^4$, a fitting formula for $\Lambda$ in terms of the mass $m$ of the boson star and of the couplings $\mu$ and $\lambda$ was obtained in Ref.~\cite{Sennett:2017etc}. In the strong coupling limit, $\lambda\gg \mu^2$, the fit reduces to~\cite{Pacilio:2020jza}
\begin{equation}
\label{fit:lambda}
        \frac{m}{m_B}=\frac{\sqrt{2}}{8\sqrt{\pi}}\left[-0.828+\frac{20.99}{\log\Lambda}-\frac{99.1}{\left(\log\Lambda\right)^2}+\frac{149.7}{\left(\log\Lambda\right)^3}\right]\,,
\end{equation}
where $m_B={\sqrt{\lambda}}/{\mu^2}$.  The above fit can be inverted to find $\Lambda=\Lambda(m/m_B)$. In this model boson stars exist only for masses smaller than a maximum one\footnote{Since $m_B$ is a free parameter of the model, boson stars can naturally populate the SSM range, depending on the coupling constants of the underlying theory.}, $m\lesssim0.06 m_B$, which implies $\Lambda\gtrsim289$. While this lower bound is much smaller than for a NS, $\Lambda$ can span many orders of magnitude as the mass deviates from its maximum value: e.g., $\Lambda\approx1.7\times10^6$ for $m/m_B=0.02$. This is again related to the strong dependence on the compactness $m/R$ [see Eq.~\eqref{Lambda}], which in this model is fitted by~\cite{Pacilio:2020jza}
\begin{equation}
    R/m\approx 7.5+48.8\left(1-\frac{m}{0.06 m_B}\right)^2\,. \label{compactnessBS}
\end{equation}
For $m/m_B=0.02$, the compactness reads $m/R\approx 0.03$ and, through Eq.~\eqref{Lambda}, $\Lambda$ is enhanced by a factor $\approx 900$ compared to the maximum mass case, $m\approx 0.06 m_B$.
Similar fits can be derived for so-called solitonic boson stars with sextic terms in the potential~\cite{Sennett:2017etc}, which support more compact configurations. Since the latter are less deformable, the minimum Love number is also smaller.

In practice, an upper bound on $\Lambda$ from a putative subsolar GW event would exclude several models of exotic compact objects~\cite{Cardoso:2017cfl,Cardoso:2019rvt,Maggio:2021ans}, and would identify the surviving ones as potential competitors to PBHs~\cite{Franciolini:2021xbq}.
Taking the above boson star case as an example, an upper bound on $\Lambda$ at the level of $\Lambda\lesssim 300$ would exclude boson stars with quartic interactions (for any choice of the couplings), but not solitonic models, while a less stringent constraint would be sufficient to exclude the minimal model with no self interactions.

\subsection{Waveform model }
We use the standard TaylorF2 waveform~\cite{Damour:2000zb} augmented with the 5PN and 6PN tidal terms in the phase~\cite{Wade:2014vqa}. 
Neglecting amplitude corrections, the TaylorF2 gravitational waveform takes the form
\begin{equation}
\tilde{h}(f)=A\, f^{-7/6}\, \exp\left[i\psi(f)\right],
\end{equation}
where the signal amplitude is proportional to $A\propto{M}_{c}^{5/6}/d_L$, the chirp mass is ${M}_{c}= (m_1 m_2)^{3/5}/(m_1+m_2)^{1/5}$, and $d_L$ is the luminosity distance between the GW detector and the binary.  

The phase is expanded 
as a power series in the parameter $x=(\pi f M)^{2/3}$. A term proportional to $x^n$ 
corresponds to the $n$-PN order of the approximation. 
In the point-particle phase, we consider standard 3.5PN circular contribution \cite{PhysRevD.44.3819,Damour:2000gg,PhysRevD.71.084008,Buonanno:2009zt}, while we also include spin effects up to 4PN order (see Refs.\cite{Franciolini:2021xbq,Favata:2021vhw} and references therein for details, e.g.~\cite{Kidder:1992fr,PhysRevD.48.1860,Kidder:1995zr,Poisson:1997ha,Mikoczi:2005dn,Gergely:1999pd,PhysRevD.74.104034,Mishra:2016whh}) assuming spins aligned with the orbital angular momentum.

We consider mergers on quasicircular orbits which, in the BH case, are overall characterized by 11 parameters (see e.g.~\cite{Maggiore:2007ulw})
\begin{equation}\label{eq:paramBBH}
{\boldsymbol \theta}
= \{m_1, m_2, d_L, \theta, \phi, \iota,  \psi, t_c, \Phi_c, \chi_{1}, \chi_{2} \}\,, 
\end{equation}
where $\chi_{1,2}$ are the aligned spin magnitudes; $\theta=\pi/2-\delta$ and $\phi$ the sky position coordinates (with $\phi$ and $\delta$ being the right ascension and declination, respectively); $\iota$ is the inclination angle of the binary with respect to the line of sight; $\psi$ the polarization angle; $t_c$ the time of coalescence; and $\Phi_c$ the phase at coalescence.

This set of parameters 
is extended when considering tidal deformation effects and potential disruption.
The 5PN and 6PN tidal corrections are added linearly to the point-particle terms as
\begin{equation}
\psi(x)=\psi_{\rm pp}(x)+\delta{\psi_{\rm tidal}}(x)\,,
\end{equation}
where the tidal contribution contains 5PN and 6PN terms~\cite{Wade:2014vqa,Lackey:2014fwa}
\begin{widetext}
\begin{equation}\label{eq:56pn}
\delta\psi_{\rm tidal}=\frac{3}{128\eta x^{5/2}}\left[ \left(-\frac{39}{2}\tilde{\Lambda}\right)x^5+ \left(-\frac{3115}{64}\tilde{\Lambda}+\frac{6595}{364}\sqrt{1-4\eta}\mbox{ }\delta\tilde{\Lambda}\right)x^6 \right],
\end{equation}
where
\begin{eqnarray}
\label{LT}
\tilde{\Lambda}&=&\frac{8}{13}\left[\left(1+7\eta-31\eta^2\right)\left(\Lambda_1+\Lambda_2\right)+\sqrt{1-4\eta}\left(1+9\eta-11\eta^2\right)\left(\Lambda_1-\Lambda_2\right)\right]\\
\label{dLT}
\delta\tilde{\Lambda}&=&\frac{1}{2}\left[\sqrt{1-4\eta}\left(1-\frac{13272}{1319}\eta+\frac{8944}{1319}\eta^2\right)\left(\Lambda_1+\Lambda_2\right) + \left(1-\frac{15910}{1319}\eta+\frac{32850}{1319}\eta^2+\frac{3380}{1319}\eta^3\right)\left(\Lambda_1-\Lambda_2\right)\right].
\end{eqnarray}
\end{widetext}
We have also introduced $\eta = m_1 m_2/(m_1+m_2)^2$, which is the symmetric mass ratio, whereas $\Lambda_1$, $\Lambda_2$ are the dimensionless tidal deformabilities of the binary components [see Eq.~\eqref{Lambda}].
Once a relation $\Lambda_i(m_i)$ is prescribed, the above waveform is valid for any compact object, including exotic ones like boson stars~\cite{Cardoso:2017cfl,Pacilio:2020jza}.

In addition to 5PN and 6PN order terms in the GW phase, in the case of SSM BNS mergers, one expects tidal disruption to take place well before the ISCO frequency, as discussed in Sec.~\ref{sec:maxfNS}.
Therefore, we assume the GW amplitude is tapered as 
$   \tilde h(f) \propto {\cal T}(f)$.
To work with dimensionless quantities, it is convenient to express the cutoff frequency as a fraction of the ISCO, i.e., $f_\text{\tiny cut} = \tilde \lambda_f f_\text{\tiny ISCO}$, in such a way that $\tilde  \lambda_f<1$ by construction. Also, we define the adimensional slope parameter as $
f_\text{\tiny slope} = \delta \tilde  \lambda_f f_\text{\tiny ISCO} $, to get 
\begin{equation}
    \tilde h(f) = 
    A  \, f^{-7/6}
    \left [
\frac{1+e^{-\tilde \lambda_f/\delta \tilde \lambda_f}}{1+e^{(f/f_\text{\tiny ISCO}-\tilde \lambda_f)
/\delta \tilde \lambda_f }}
\right ]
\, \exp\left[i\psi(f)\right].
\end{equation}
Therefore, in case one wants to test the presence of tidal effects, the parameters to add to the list in Eq.~\eqref{eq:paramBBH} are 
\begin{equation}\label{eq:tidaldefparam}
{\boldsymbol \theta}_T
= \{\tilde \Lambda, 
\delta \tilde \Lambda ,
\tilde \lambda_f ,\delta \tilde \lambda_f\}\,,
\end{equation}
and the waveform model will overall contain $15$ parameters.

We stress that ours is a simplified model aimed at capturing the salient features of tidal disruption in the waveform. Improved waveform modeling SSM BNS, based on numerical simulations, would be of utmost importance to support strong claims on the nature of eventual SSM detections.

\section{Results}

In this section we will discuss the tests performed using our models in various settings presented in Sec.~\ref{subsec:tests}. We have performed parameter estimation through both a Fisher information matrix analysis using {\tt GWFast}~\cite{Iacovelli:2022mbg,Iacovelli:2022bbs} and a more computationally demanding Bayesian inference using the public software {\tt BILBY}~\cite{BILBY}. Both methods are standard and reviewed in Appendix~\ref{app:methods}. 

We tested our Bayesian analysis by reproducing the results of Ref.~\cite{Prunier:2023cyv} using the same settings (i.e., neglecting tidal effects and including spin precession) and after an injection compatible with the subthreshold event SSM200308.

Then, in Sec.~\ref{subsec:SSMevent} we consider a SSM200308-like event, but this time including tidal effects in the analysis and considering different detector sensitivities. We have compared the results of the Fisher matrix with those of the Bayesian inference, finding consistent results even for low SNR. This justifies the use of the Fisher matrix results to explore the full parameter space, as done in Sec.~\ref{subsec:paramspace}.

\subsection{Tests and diagnostics} \label{subsec:tests}

We forecast observational prospects to test the nature of SSM detections by injecting different signals and considering two distinct scenarios, discussed below.

\subsubsection{BNS case}
We assume that the GW signal comes from a SSM BNS. In this case, we inject the values of $\tilde \Lambda$ and $\tilde \delta \Lambda$ obtained from the tabulated values of $\Lambda_i(m_i)$ taking SLy4 EoS as representative case (as previously discussed, in this regime the EoS uncertainty is negligible). Furthermore, we will inject the tapering coefficients $\tilde \lambda_f = f_\text{\tiny RO}/f_\text{\tiny ISCO} $ from Eqs.~\eqref{eq:fisco} and \eqref{eq:RO}, while we set $\delta \tilde \lambda_f = \tilde \lambda_f/6$.

Assessing or ruling out the PBH nature of a SSM event is conveniently done through a hierarchical procedure, see~\cite{Franciolini:2021xbq} for details.
First of all, for a given event, we should make sure that the precision on either mass measurement is sufficient to claim the event had at least one SSM object. At $3\sigma$ confidence level, we therefore require\footnote{A similar criterion was proposed in Ref.~\cite{Mancarella:2023ehn} to define the 
inference horizon redshift of next-generation detectors.}
\begin{equation}
    m_i + 3 \Delta m_i < M_\odot\,,  \label{cond_mass}
\end{equation}
for $i=1$ or $i=2$, where $\Delta m_i$ is the standard deviation on the $i$th binary component mass.

Then, to exclude the PBH nature of a SSM binary, we require the measured values of $\tilde \Lambda$ to be incompatible with zero, and/or values of $\lambda_f$ incompatible with unity.  This translates to
\begin{eqnarray}
     \tilde{\Lambda}-3\Delta{\tilde\Lambda}&>&0 \,,\quad {\rm and/or} \label{cond_Lambda}\\
     \tilde\lambda_f+3\Delta{\tilde\lambda_f}&<&1 \label{cond_lambda}
\end{eqnarray}
again for a test at $3\sigma$ confidence level.
%

\subsubsection{BPBH case}
We assume the GW signal comes from a SSM PBH binary.
In this case, we inject a signal with no tidal deformability effects ($\tilde \Lambda=\delta\tilde\Lambda=0$), which lasts at least until the ISCO frequency ($\tilde\lambda_f=1$). In practice, for the SSM binaries we are interested in, the ISCO frequency is too high to be detectable by ground-based detectors, see Eq.~\eqref{eq:fisco}. Therefore, in this case we can directly ignore the tapering parameters $\tilde\lambda_f$ and $\delta\tilde\lambda_f$ since the signal is effectively independent of them.

Also in this case the condition~\eqref{cond_mass} must hold to ensure that at least one of the binary components is subsolar. However, this scenario would only allow setting upper bounds on the tidal deformability, potentially excluding expected values for BNS (e.g., the prediction in Eq.~\eqref{Sly4} for a given mass) or for more exotic models.
%

\subsection{SSM200308-like events} \label{subsec:SSMevent}
Let us start by evaluating the performance of current and future detectors when inferring the properties of signals which are similar to subthreshold candidate events reported by LVK \cite{LIGOScientific:2022hai}. 
To do so, we inject a system with similar properties to the candidate event SSM200308 with $m_1=0.62M_\odot$, $m_2=0.27 M_\odot$.
We also inject negligible spins, because this hypothesis is compatible with theoretical expectations for both the PBH case when formed in radiation dominated universe~\cite{DeLuca:2019buf,Mirbabayi:2019uph}\footnote{
It is possible to form PBHs in different scenarios, such as from the assembly of matterlike objects (i.e. particles, Q-balls, oscillons, etc.), domain walls and heavy quarks of a confining gauge theory. This can lead to different predictions for the PBH spin at formation~\cite{Harada:2017fjm,Cotner:2019ykd,Flores:2021tmc,Dvali:2021byy,Eroshenko:2021sez}.
For instance, during an early matter-dominated phase PBH could develop initial large spins~\cite{Harada:2017fjm,DeLuca:2021pls}. } and the NS case. 
We checked that our results do not qualitatively depend on the value the injected spins. For simplicity, we neglect any spin misalignment, even though Ref.~\cite{Prunier:2023cyv} reported evidence for spin precession in this particular event.
We adopt the strategy discussed in Sec.~\ref{subsec:tests}, assuming the event was either a BNS 
(injecting $\tilde \Lambda  = 1.5 \cdot 10^5 $, 
$\delta \tilde \Lambda =  4.9 \cdot 10^4 $, and $\tilde \lambda_f = 0.075 $), or a BPBH
(injecting $\tilde \Lambda = \delta \tilde \Lambda = 0 $ and $\tilde \lambda_f = 1$).
We consider a signal in the frequency band $f\in[10\,{\rm Hz},f_{\rm max}]$, with $f_{\rm max}=2048\,{\rm Hz}$ for O3-O5 and $f_{\rm max}=4096\,{\rm Hz}$ for ET/CE.

The results of a Fisher analysis are reported in Table~\ref{tab:res} for the two cases. 
The uncertainties on both masses computed assuming O3 sensitivity are in very good agreement  with the parameter estimation performed in Ref.~\cite{Prunier:2023cyv} within the BPBH assumption, when accounting for the different range of frequencies adopted in their analysis of the data. Uncertainties are larger if one assumes a BNS signal, due to its lower cutoff frequency and the smaller number of cycles in the sensitive band.

{
\renewcommand{\arraystretch}{1.4}
\setlength{\tabcolsep}{4pt}
\begin{table}[t!]
\begin{tabularx}{\columnwidth}{|X|c|c|c|c|}
\hline
\hline
Network  & LVK O3 & LVK O4 & LVK O5 & ET+2CE \\
\hline
\hline
\multicolumn{5} {|c|} {BNS SSM200308
($\tilde \Lambda  = 1.5 \cdot 10^5,  
\delta \tilde \Lambda =  4.9 \cdot 10^4, 
\tilde \lambda_f = 0.075 $)} 
\\
\hline
SNR & 7.90 & 12.8 & 22.4 & 398 \\
\hline
$\Delta m_1/ m_1$   &
0.47 & 0.22 & 0.082&  0.0017\\
\hline
$\Delta m_2/ m_2$   & 
0.39  & 0.19  & 0.070& 0.0015 \\
\hline
$\Delta \tilde \Lambda / \tilde \Lambda$    & 
0.86  & 0.66  & 0.55 & 0.047 \\
\hline
$\Delta \tilde \lambda_f / \tilde \lambda_f $    & 
 0.38 & 0.24  & 0.13  & 0.015  \\
\hline
\hline
\multicolumn{5} {|c|} {BPBH SSM200308 
($\tilde \Lambda = \delta \tilde \Lambda = 0, 
\tilde \lambda_f = 1$)} 
\\
\hline
SNR & 8.76 & 14.6 & 24.8 & 430 \\
\hline
$\Delta m_1/ m_1$   &
 0.21 &  0.14 & 0.053 & $6.4\cdot 10^{-3}$   \\
 \hline
$\Delta m_2/ m_2$   &  
0.18 & 0.12 & 0.046 & $5.5\cdot 10^{-3}$  \\
\hline
$ \Delta \tilde \Lambda $    & 
 $1.9 \cdot 10^4$ & $1.3 \cdot 10^4$  & $7.8 \cdot 10^3$ & $7.7 \cdot 10^2$  \\
\hline
\hline
\end{tabularx}
\caption{Fisher parameter estimation uncertainties with current and future GW experiments. We inject a system with similar properties to the subthreshold event SSM200308 with $m_1=0.62M_\odot$ and $m_2=0.27 M_\odot$, assuming the object was either a BNS (top rows) or a BPBH (bottom rows).  
}
\label{tab:res}
\end{table}
}

\begin{figure*}[!t]
\centering
\includegraphics[scale = 0.44]{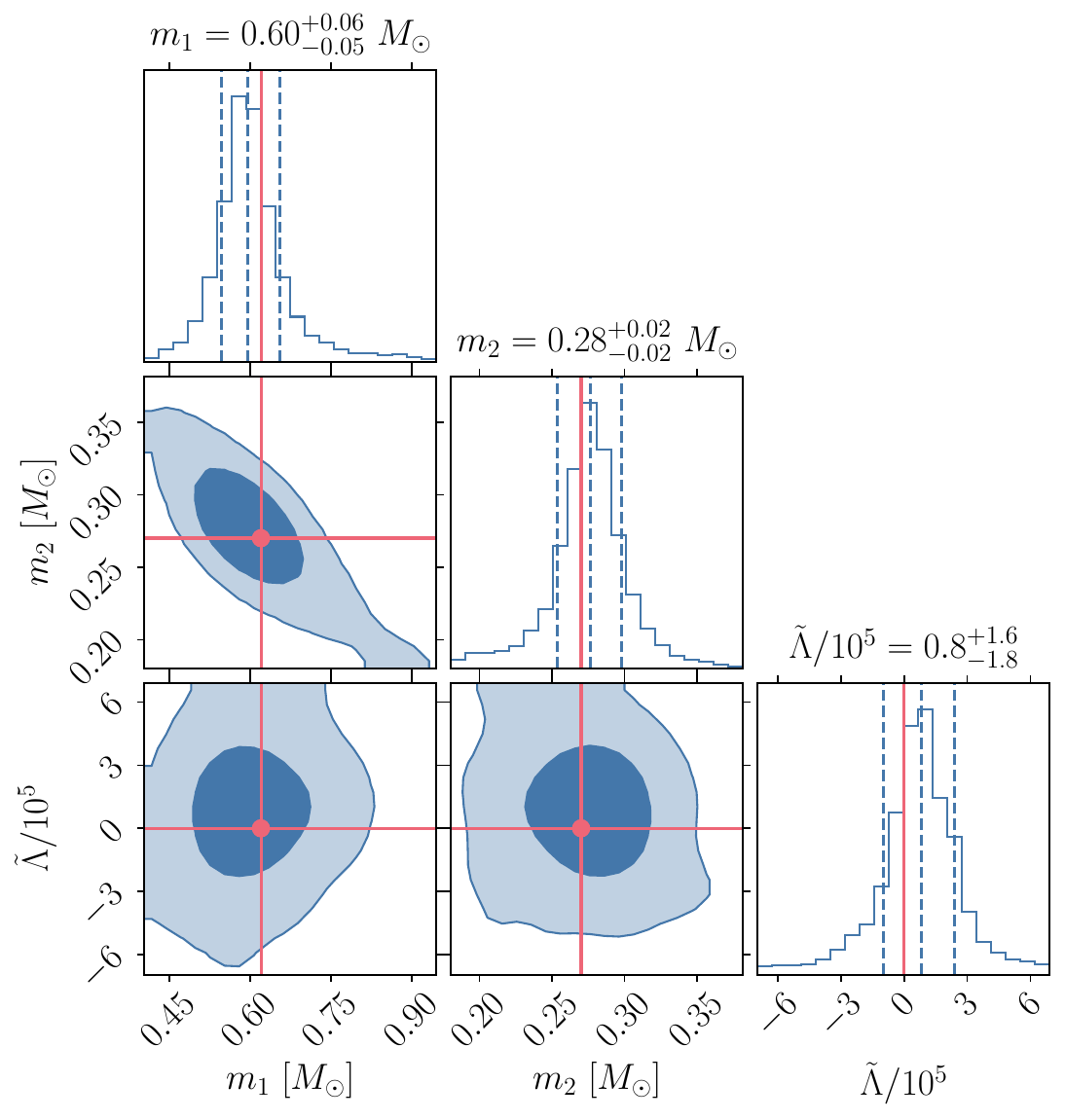} 
\includegraphics[scale = 0.44]{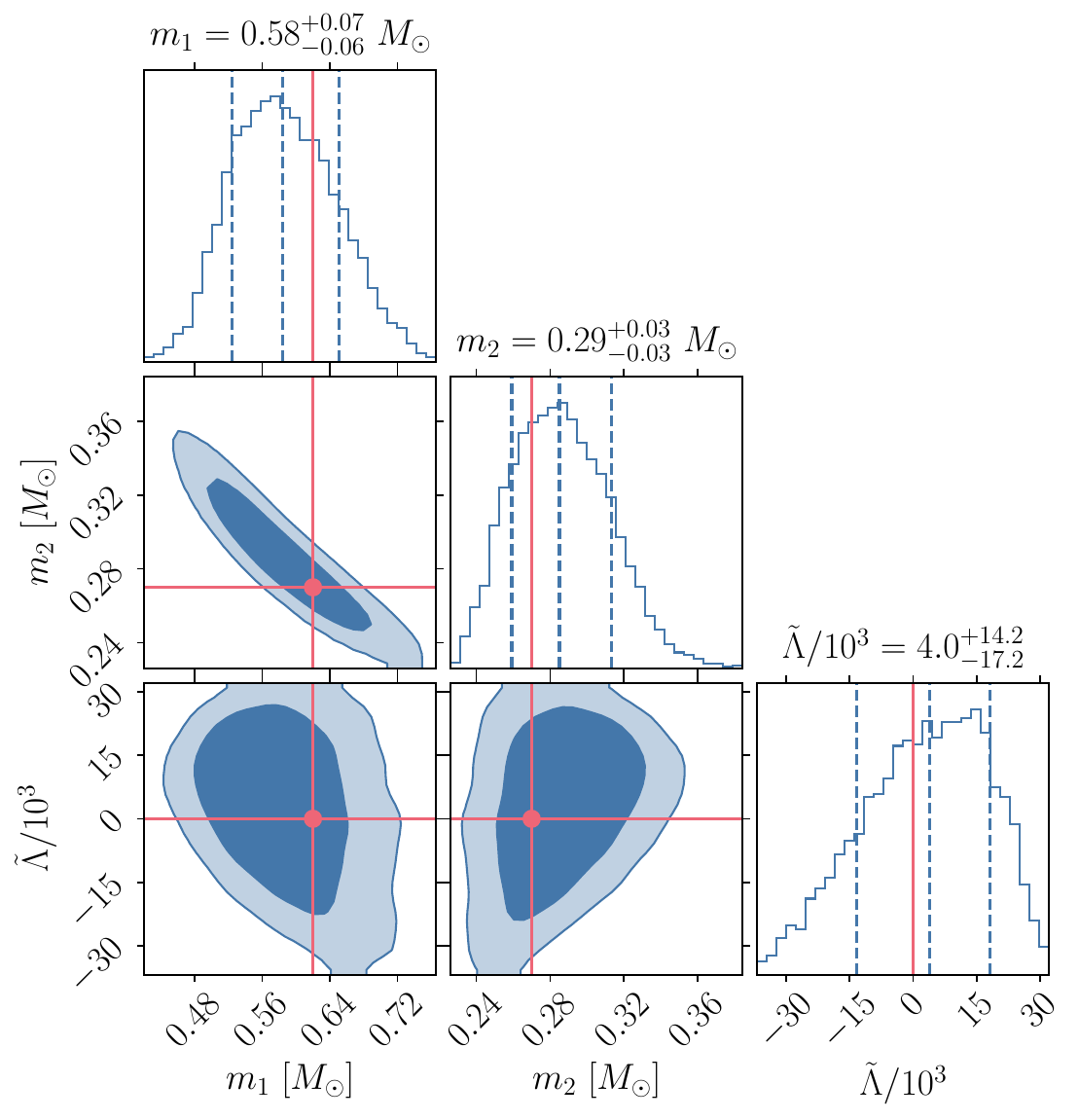}
\includegraphics[scale = 0.44]{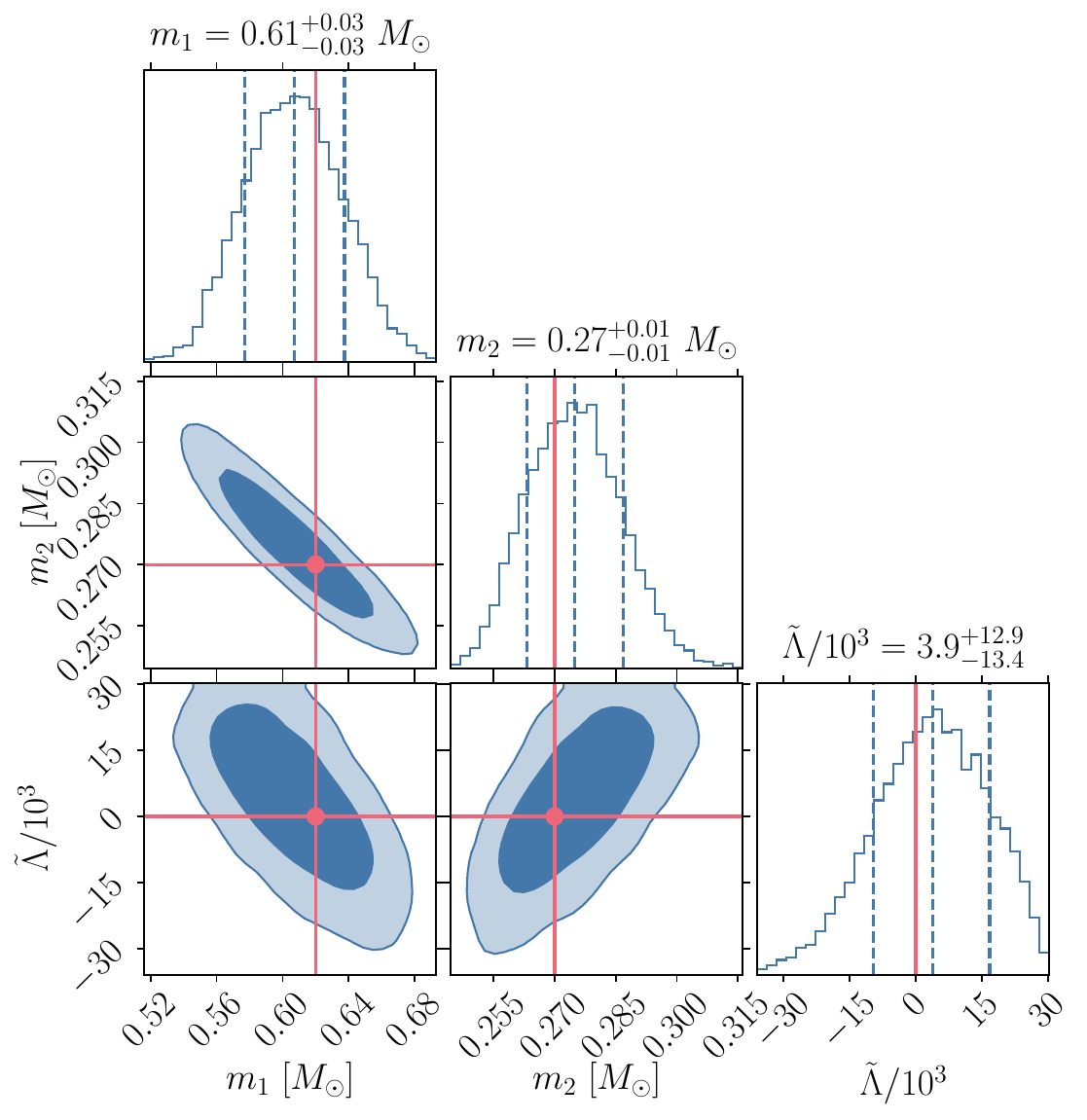} 
\includegraphics[scale = 0.44]{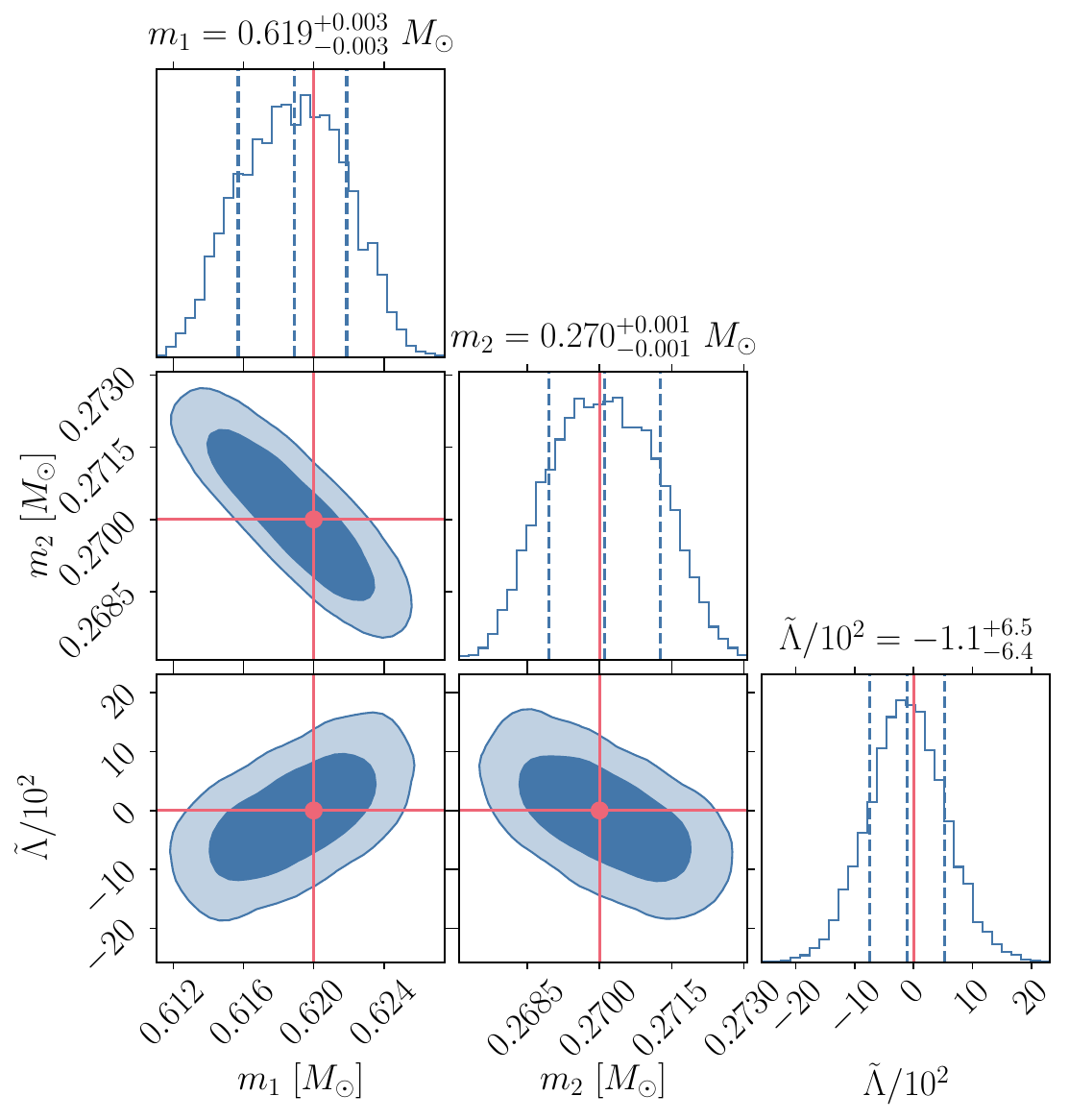}
\caption{Posterior distribution of the source frame masses $(m_1,m_2)$ and the $\tilde{\Lambda}$ parameter for respectively O3 (top left), O4 (top right), O5 (bottom left) and ET+2CE (bottom right). The levels of the 2D joint distributions indicate respectively the 68$\%$ and $95\%$ credible levels, while the dashed lines on the 1D distributions are the 1$\sigma$ intervals with respect to the median. The orange lines represent the injected values.}
\label{fig:corner}
\end{figure*}

The Fisher-matrix results are corroborated by a full-fledged Bayesian analysis, which provides consistent $1\sigma$ errors as those shown in Table~\ref{tab:res}. An example is presented in Fig.~\ref{fig:corner} for the BPBH injection. 
We show a subset of the posterior distributions for O3, O4, O5, ET+2CE in the four panels, respectively.

Interestingly, while the Bayesian analysis provides larger errors on $\tilde\Lambda$ for O3 (${\rm SNR}\approx8$), showing the limitations of the Fisher matrix estimates for low SNR signals, already starting from O4 (${\rm SNR}\approx13$) both errors are in very good agreement with each other. The Fisher-matrix errors on the masses are overestimated for the relatively low SNRs in O3 and O4, but also in this case the agreement is very good at larger SNR, as expected.
Although not shown, we found the same level of agreement also for the other waveform parameters.

Let us now discuss these forecasts in more detail. First of all we notice that, in the BNS case, measurements errors in O3 are such that the primary mass does not satisfy the condition in Eq.~\eqref{cond_mass}, whereas it marginally satisfies it in O4, and both events are confidently measured as subsolar only starting from O5.
As discussed, errors are smaller in the BPBH case and indeed $m_1$ satisfies Eq.~\eqref{cond_mass} already in O3~\cite{Prunier:2023cyv}. The lighter object is always well within the SSM range, even when we augment the waveform with tidal effects (hence increasing the errors).

Then, we notice that in the BNS the errors on the tidal deformability are relatively large and Eq.~\eqref{cond_Lambda} is never satisfied for LVK. Only third-generation detectors~\cite{Branchesi:2023mws} will allow to exclude $\tilde\Lambda=0$ at least at $3\sigma$ level (at $>5\sigma$ level, in fact). However, the cutoff frequency for tidal disruption is measured more accurately. Indeed, the condition in Eq.~\eqref{cond_lambda} is satisfied already by O4, which actually reaches $4\sigma$ confidence level. This shows that even current GW observations can confidently rule out the PBH nature of a putative SSM200308-like event by measuring tidal effects. Starting from O4, the same conclusion can be reached at $5\sigma$ and higher confidence level. 

As previously discussed, in the BPBH case our agnostic test can at most put an upper bound on $\tilde\Lambda$, ruling out competitive models that predict larger tidal deformability.
For example, Eq.~\eqref{Sly4} predicts that, to be compatible with the BNS hypothesis, a SSM200308-like event should have $\tilde\Lambda\approx 1.5\cdot 10^5$, which is several sigmas in tension with the errors shown in the bottom rows of Table~\ref{tab:res} already for O3.
Therefore, if a SSM200308 was a BPBH, we could confidently exclude the competitive BNS hypothesis. 

On the other hand, this might not be the case for more exotic (and more compact) models. The example of boson stars with quartic interactions shows that $\Lambda_i\gtrsim289$ when $m_i\lesssim 0.06 m_B$~\cite{Pacilio:2020jza}. For equal-mass binaries, this implies that $\tilde\Lambda\gtrsim289$ so, depending on the value of $m_B$, it could be small enough to stay within errors. Specifically, for $m_1=m_2=0.62 M_\odot$, solutions exist only for $m_B\gtrsim10.3 M_\odot$ and one get $\tilde\Lambda>3 \cdot 10^4$ when $m_B\gtrsim 15M_\odot$. In other words, for SSM200308-like errors in O3, also the quartic boson star binary would be excluded at high significance level except for the small range  $10.3\lesssim m_B/M_\odot\lesssim 15 $, in which $\tilde\Lambda$ is sufficiently small.

Furthermore, in this specific model the constraints becomes even more stringent in the unequal-mass case, because even if the parameter $m_B$ is tuned to minimize $\Lambda_1\approx 289$, the lighter companion is necessarily less compact and would have bigger $\Lambda_2$, which dominates the values of $\tilde\Lambda$. Using the masses estimated for SSM200308,  it turns out that $\tilde\Lambda\gtrsim 4\cdot 10^4$, being thus excluded at $2\sigma$ already in O3.

In other words, depending on the mass ratio the tidal deformability parameter $\tilde\Lambda$ can be mostly accounted for by the deformability of the lighter companion which, in a specific model, is also the least compact (and hence more deformable) one. This makes the upper bounds on $\tilde\Lambda$ derived for an unequal-mass binary more effective to rule out specific models.

\begin{figure*}[t!]
\centering
\includegraphics[width=0.32\textwidth]{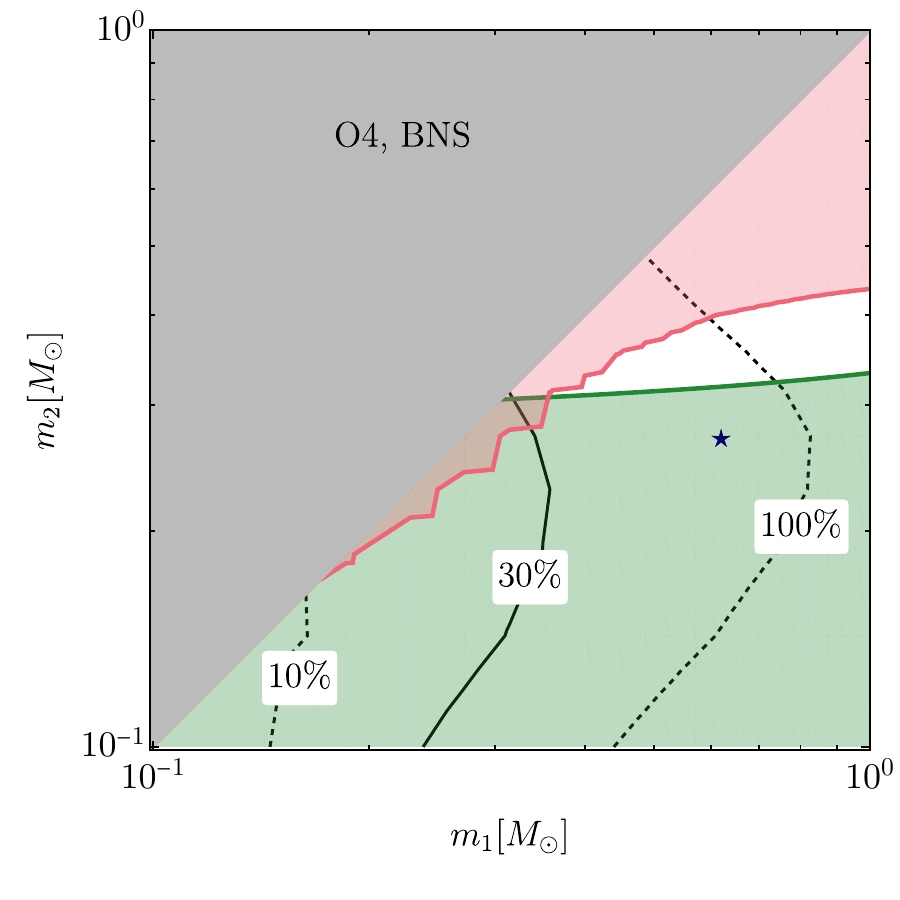}
\includegraphics[width=0.32\textwidth]{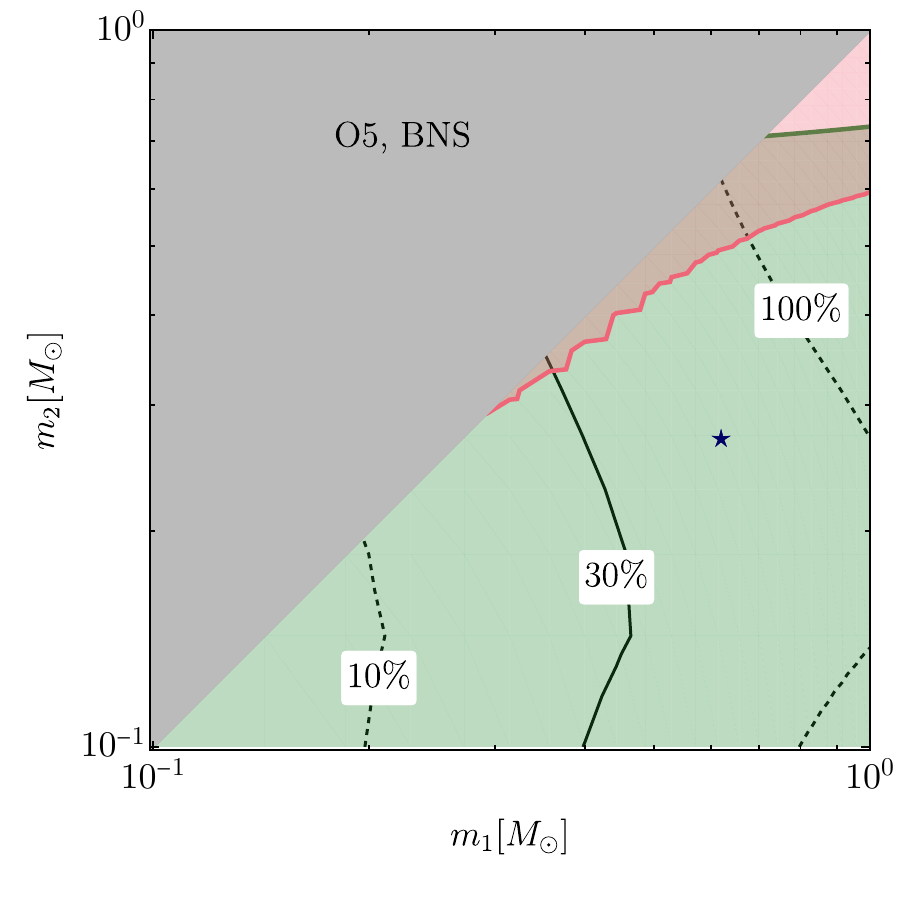}
\includegraphics[width=0.32\textwidth]{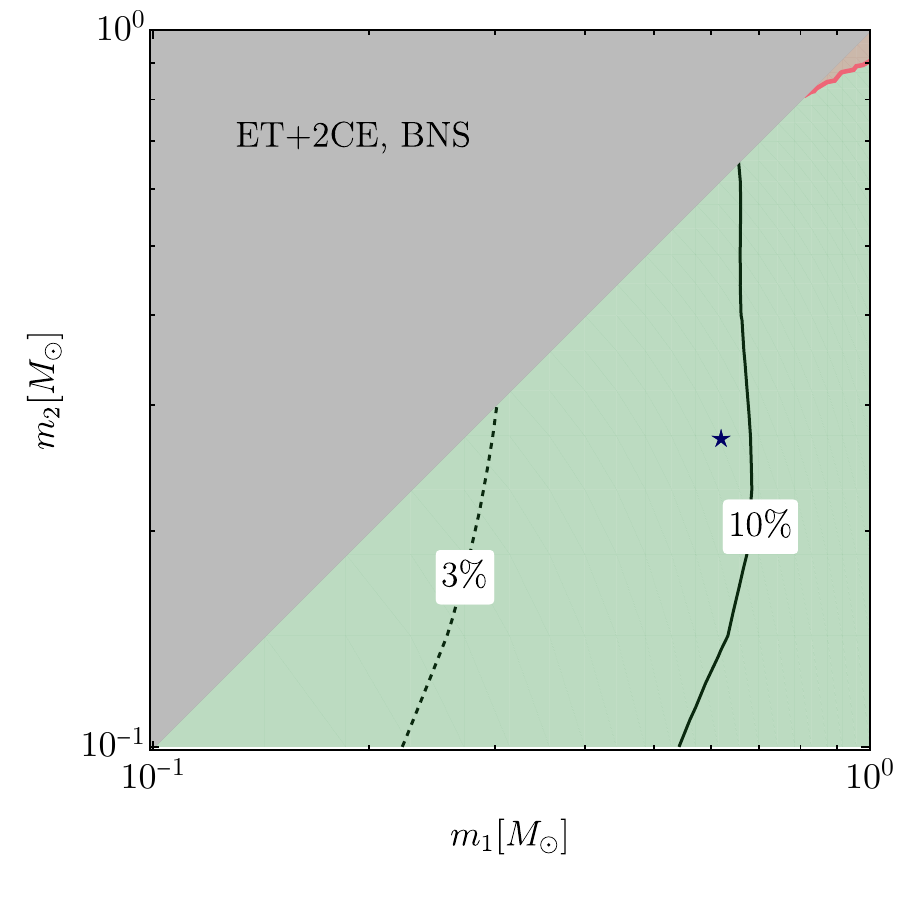}
\caption{ 
We show in green the parameter space where $\tilde \lambda_f$ is constrained to be smaller than 1 at $3\sigma$ C.L., assuming the injected signal is a BNS. 
Below the red shaded region, at least one of the binary components is constrained to be subsolar with $3\sigma$ confidence. 
The black lines indicate contour levels for relative precision on $\tilde \Lambda$.
In the left, center, and right panels, we consider O4, O5, and ET+2CE detector configurations. 
To facilitate comparison between detectors, we placed the optimally oriented binary at a distance such that SNR = 12 with O4 sensitivity in all panels. 
In the right panel, almost in all parameter space, one can confidently observe at least one SSM component, due to the large precision precision achieved by next-generation detectors. 
The blue star indicates the best fit masses for a SSM200308-like event.
}\label{fig:m1m2bns}
\end{figure*}

\subsection{Exploring the SSM parameter space: Generic forecasts}
\label{subsec:paramspace}

\begin{figure*}[t!]
\centering
\includegraphics[width=0.32\textwidth]{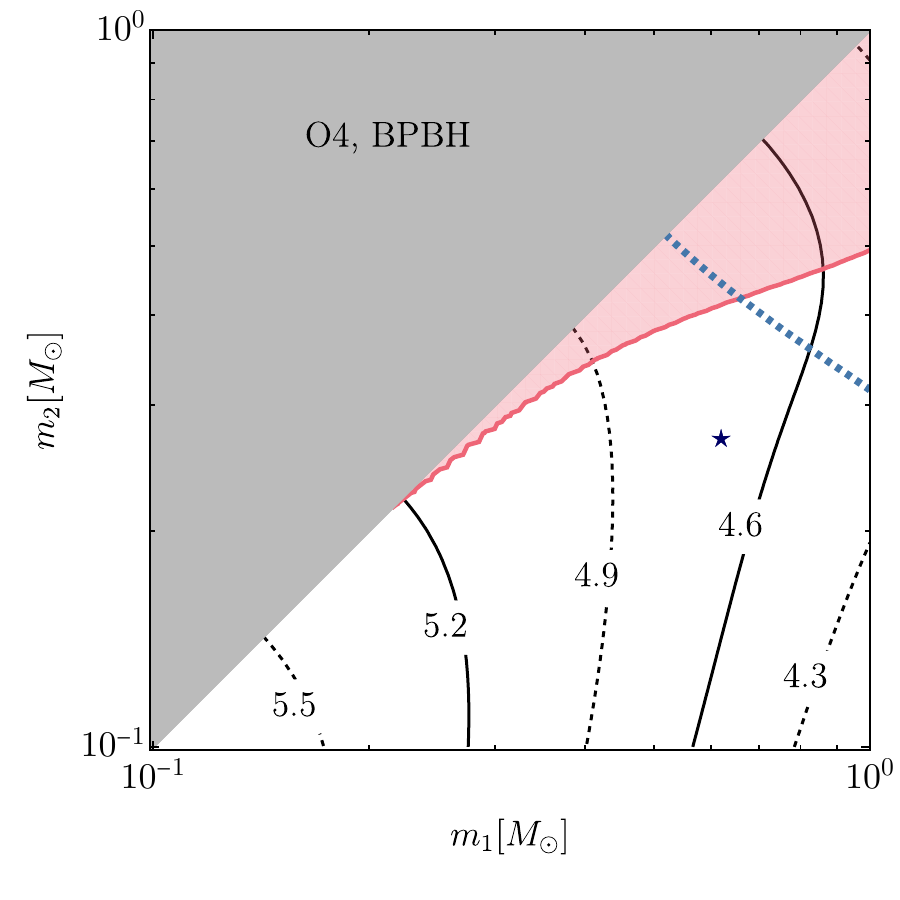}
\includegraphics[width=0.32\textwidth]{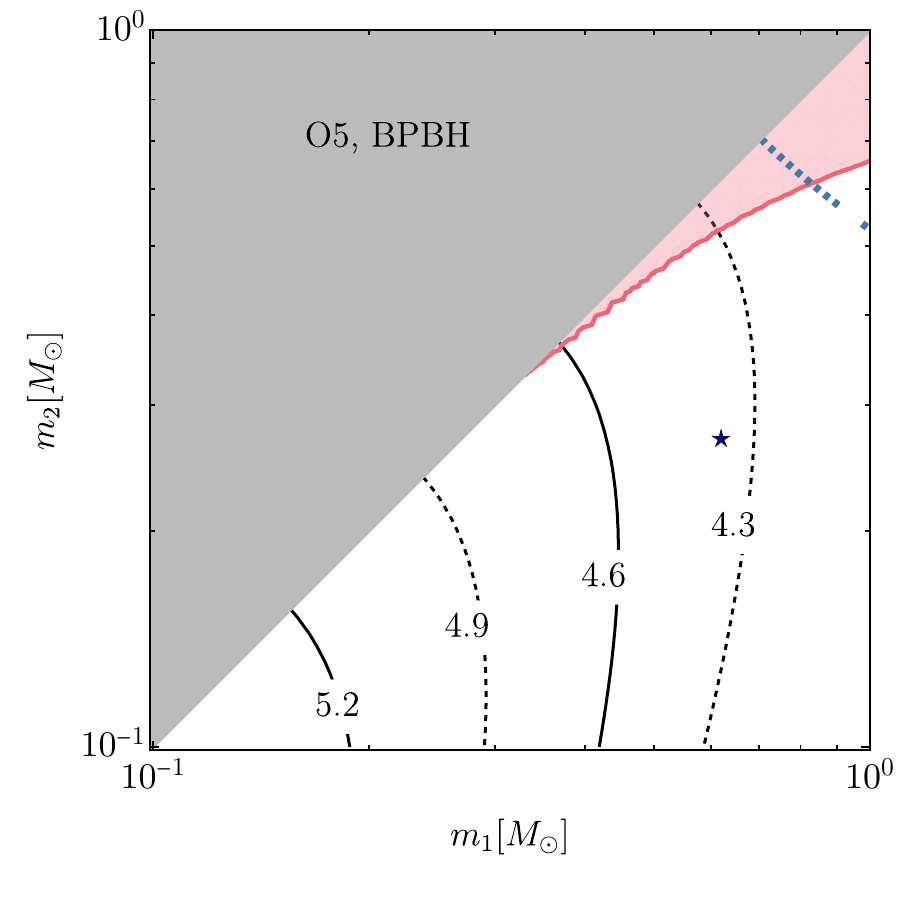}
\includegraphics[width=0.32\textwidth]{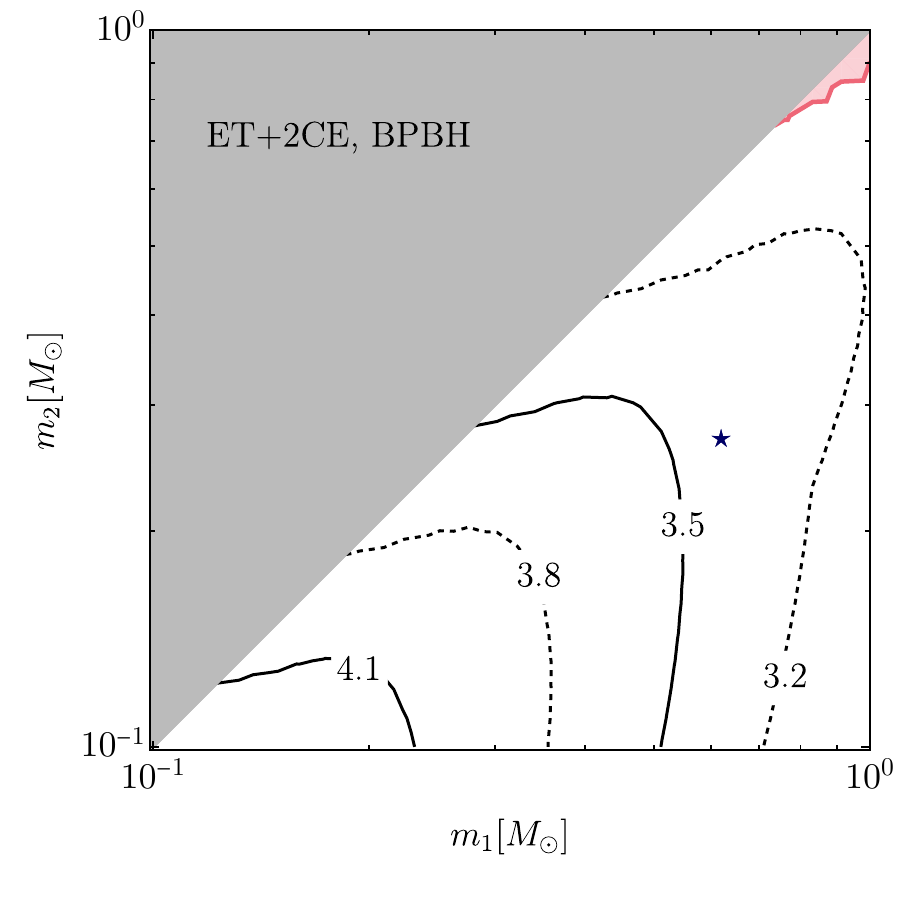}
\includegraphics[width=0.32\textwidth]{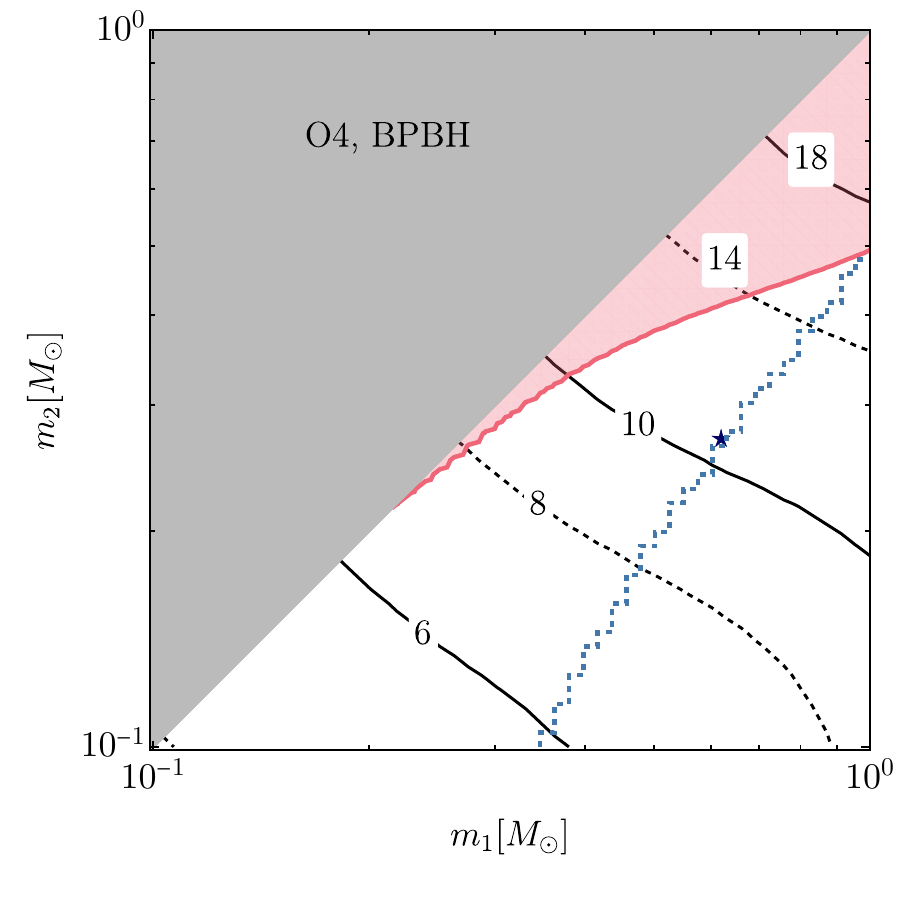}
\includegraphics[width=0.32\textwidth]{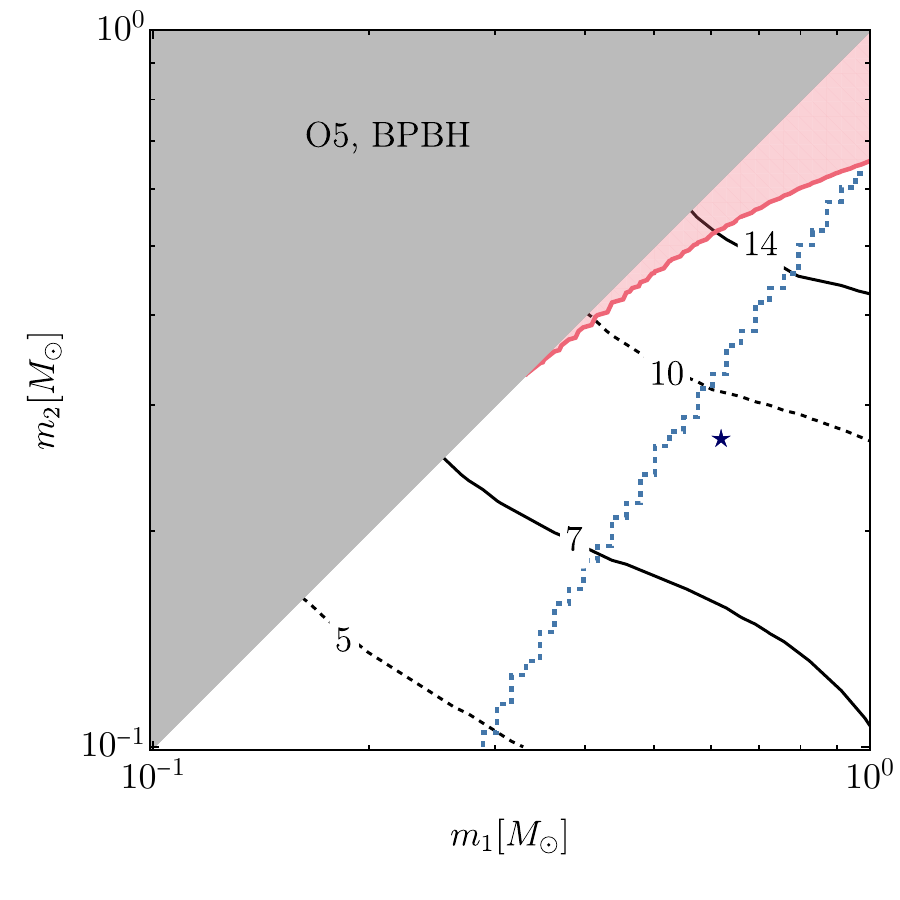}
\includegraphics[width=0.32\textwidth]{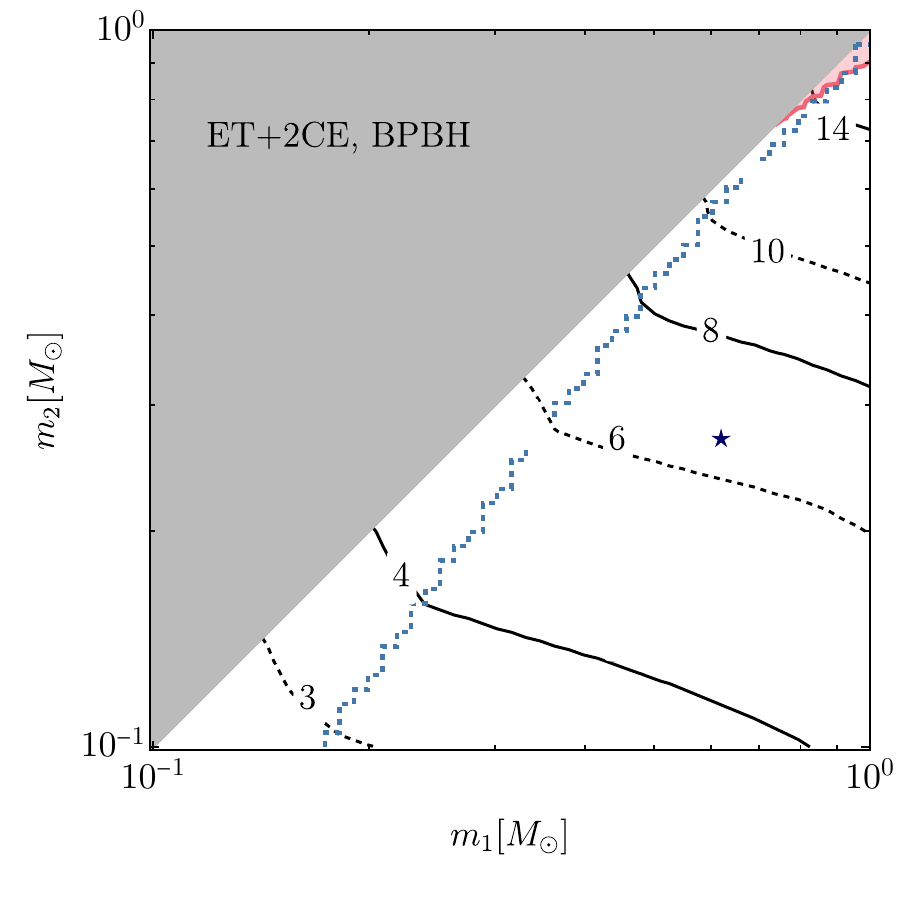}
\caption{ 
Same as Fig.~\ref{fig:m1m2bns} but injecting a PBH binary. 
{\it Top panels:}
the black contour lines show the ($\log_{10}$)
upper bound on the effective tidal deformability 5PN term, 
$\tilde \Lambda$,  at $3\sigma$ C.L. 
Below the dashed diagonal blue lines the tidal deformability for BNS (assuming SLy4 Eos) would be ruled out at more than $3\sigma$ C.L.
{\it Bottom panels:}
the contour lines indicate values of $m_B/M_\odot$ above which the tidal deformability would be incompatible with future upper bounds (at $3 \sigma$ C.L.).
We assume a boson star model with quartic potential where $m_B={\sqrt{\lambda}}/{\mu^2}$ and the compact object tidal deformability is given in Eq.~\eqref{fit:lambda}.
Below the blue diagonal line, the condition on the maximum boson-star mass cannot be satisfied while simultaneously providing a compact enough solution compatible with the upper bounds shown in the contour plot. Therefore, in this region the boson-star model can be entirely ruled out.
}\label{fig:planem1m2}
\end{figure*}

After having focused on a SSM200308-like event, we now take a broader perspective and explore the entire SSM parameter space of interest for ground-based detectors.

In Fig.~\ref{fig:m1m2bns}, we report the relative precision achieved when measuring both $\tilde \Lambda$ and $\tilde \lambda_f$ with future O4, O5, and ET+2CE detectors in the SSM range. We scan the parameter space where both masses are in the range $(m_1,m_2) \in [0.1,1]$. We assume the binary is optimally oriented at a distance corresponding to the threshold for detection with O4 sensitivity.\footnote{Note that this implies that the luminosity distance varies depending on the masses, and it decreases in the bottom leftmost part of the diagrams. For example, when $m_1=m_2=0.1 M_\odot$ we have only $d_L\approx 50\,{\rm Mpc}$ for a binary at the detection threshold in O4.}
In order to explore the entire parameter space, here we only resort to a Fisher analysis, which proved to give reliable results for the case study of the previous section. The errors obtained would scale inversely with SNR for louder signals. 

We shade in green the region where the GW cutoff frequency would be incompatible with $f_\text{\tiny ISCO}$ at $3\sigma$ C.L., see Eq.~\eqref{cond_lambda}. Furthermore, in the same panel, we also shade in red the region where precision on either mass would not be sufficient to claim the event had at least one of the objects in the SSM range, i.e. when Eq.~\eqref{cond_mass} is not satisfied. In the same plot, we also report contour lines indicating the relative uncertainty on $\tilde \Lambda$.

As we can see, during the O4 run one would be able to confirm a SSM detection is consistent with being a BNS if the secondary mass is $m_2\lesssim  0.3 M_\odot$, for all mass ratios considered. 
This information would mostly come from the cutoff frequency of the waveform, while uncertainties on $\tilde \Lambda$ would remain large. 
As already pointed out, this happens because, due to the small Roche overflow frequency characterizing light BNS, the inspiral would be stopped much before $f_\text{\tiny ISCO}$, therefore suppressing the information contained in phase evolution from high PN orders.
In the center and right panels of Fig.~\ref{fig:m1m2bns}, we report the same results for the future LVK and next-generation experiments. 
As one can notice, large improvement on the measurability of $\tilde \lambda_f$ will be achieved within O5, covering most of the parameter space (and, in fact, in the top rightmost part of the parameter space the limiting factor becomes the accurate measurement of subsolar masses). However, even in this case the limited precision on $\tilde \Lambda$ would still not allow to measure nonzero values (at $3\sigma$ C.L.) for $m_1\gtrsim 0.4 M_\odot$.
Finally, exquisite precision will be reached with a network of detectors which includes ET and two CE observatories. We chose this configuration as the most optimistic scenario, but actually even a single third-generation detector would dramatically improve the statistical significance of the proposed tests.

In Fig.~\ref{fig:planem1m2}, we analyse the opposite case in which the injected signal is sourced by a PBH binary. In this case, one can only set upper bounds on the tidal deformability parameters, being the 5PN term the most stringent one. 
We see that, for subsolar masses $m_i\in[0.1,1] M_\odot$, upper bounds on the effective tidal deformability would fall in the range $\tilde \Lambda \lesssim 10^{(-5\div -4)}$, with a nearly linear scaling with $1/m_1$, for threshold events with ${\rm SNR}= 12$ in O4. This ballpark constraint would already be enough to rule out with large significance the tidal deformability expected for SSM BNS, see Eq.~\eqref{Sly4}. Indeed, the region below the blue dotted curve in the top panels of Fig.~\ref{fig:planem1m2} is where the BNS hypothesis can be ruled out at more than $3\sigma$ confidence level. As evident, starting from O5 this region covers basically the entire parameter space.

On the flip side, within this agnostic test it would be harder to exclude more exotic models with smaller tidal deformability that would not be in tension with the upper bounds on $\tilde\Lambda$. 
In this case, a detailed model selection between the PBH hypothesis and a given exotic compact object hypothesis would be necessary and should be performed on a case-by-case basis.
In the bottom panels of Fig.~\ref{fig:planem1m2} we perform this analysis for the aforementioned model of boson stars with large quartic interactions, in which case the bounds on $\tilde\Lambda$ can be translated into bounds on the sole model parameter, $m_B$. The contour lines indicate values of $m_B$ which would provide a $\tilde\Lambda$ [see Eq.~\eqref{fit:lambda}] incompatible with future upper bounds. Furthermore, in this model boson stars exist only when $m_i<0.06\, m_B$, providing a lower bound on $m_B$ in the $(m_1,m_2)$ plane. This is shown in the bottom panels of Fig.~\ref{fig:planem1m2} by the blue diagonal curve, below which the upper bound on $m_B$ would be lower than the lower bound required by existence of the solution. In other words, below the blue diagonal curve the model can be completely ruled out.

Note that, in this specific models, the bounds on $m_B$ do not improve dramatically for future detectors, since $m_B$ depends logarithmically on $\tilde\Lambda$, see Eq.~\eqref{fit:lambda}.

\section{Discussion and conclusions}\label{sec:conclusions}
Overall, our conclusions are quite positive. Subsolar candidate events which were on the verge of detectability in LVK O3 run could be detectable starting from O4. Even more importantly, our results show that not only for such events can the masses be measured sufficiently well to confirm the detection of a subsolar object, but also the tidal effects can be measured with sufficient accuracy, at least to conclusively confront the PBH hypothesis against the subsolar NS one, or even against more exotic hypotheses.

There are various ways in which a given event can be in tension with either the PBH or the NS hypothesis. Let us list some examples:
(i) a subsolar event with a robust measurement of nonzero tidal deformability cannot be a PBH binary, unless one invokes strong beyond-Standard Model environmental effects such as bosonic condensates\footnote{Ref.~\cite{DeLuca:2021ite} shows that bosonic condensates around BHs are compact and dense enough to provide a significant tidal contribution to the GW signal before being tidally destroyed. On the other hand, ordinary dark-matter haloes, even when accounting for steep density profiles due to accretion onto the BH, are too dilute and are destroyed much before their tidal deformability can significantly contribute to the GW phase~\cite{ValerioPrivate}.
Indeed, for a dark-matter density $\rho_\text{\tiny DM}\sim r^{-9/4}$~\cite{Bertschinger:1985pd,Adamek:2019gns}, one can estimate the Roche radius as $r_\text{\tiny RO}\sim [3 m/(4\pi \rho_\text{\tiny DM})]^{1/3}$, where $m$ is the companion mass. Equating this to the binary semi-major axis gives an estimate of the Roche frequency, $f_\text{\tiny RO}\sim 10^{-14}\,{\rm Hz}$, well below any detector bandwidth.} around PBHs~\cite{DeLuca:2021ite,DeLuca:2022xlz};
(ii) if large spins are detected this would be in tension at least with the PBH formation scenario~\cite{DeLuca:2019buf,Mirbabayi:2019uph} in which PBHs are created from the collapse of large overdensities in the radiation dominated early universe. Note also that accretion in the early universe should be negligible in this mass range~\cite{DeLuca:2020qqa,DeLuca:2020bjf}, thus not being able to spin up PBHs after formation; 
(iii) in case a nonzero tidal deformability is also detected, large spins would also be in tension with the NS hypothesis, since the spin of a NS is expected to be at most moderate and typically negligible.
The subsolar range is particularly interesting in this context because the interpretation would not be contaminated by possible hierarchical mergers~\cite{Gerosa:2017kvu}, so the binary spins are likely the natal ones.

On the other hand, even just an upper bound on the tidal deformability which is compatible with zero would support the PBH hypothesis and, as our results show, can easily be in tension with the subsolar NS one.
Should this be the case in future events, one would need to invoke either more exotic models of compact objects (also possibly of primordial origin) with smaller tidal deformability, or
subsolar BHs of nonprimordial origin, for example formed from WD or NS transmutation triggered by asymmetric or nonannihilating dark matter accretion~\cite{Takhistov:2017bpt,Kouvaris:2018wnh,Takhistov:2020vxs,Dasgupta:2020mqg,Bhattacharya:2023stq,Chakraborty:2024eyx}.
Assessing which hypothesis is then favored by the data would become imperative and would most likely require a (model-dependent) Bayesian selection.
In any case, an event of this kind would be a smoking gun for new physics, demanding a thorough and highly accurate scrutiny.

We conclude by discussing a few caveats that should be addressed in the future to extend our analysis.

Our results are based on the analytical TaylorF2 waveform approximation, possibly augmented with the inclusion of tidal terms and a tapering function suppressing the signal beyond the Roche overflow frequency. While this implementation of tidal effects is clearly simplistic, it should capture the salient features of complicated tidal disruption effects in the waveform while allowing us to devise a test that seeks to be as agnostic as possible.

Waveforms calibrated using numerical relativity simulations for SSM BNSs are not available at the time of writing, and this limitation is even more sever for simulations involving exotic compact objects. Our results underscore the importance of such simulations to improve waveform models in a seldom explored parameter space.
Recently, Ref.~\cite{Markin:2023fxx} performed a numerical simulation of a NS-SSMBH system with masses $1.4 M_\odot$ and $0.5 M_\odot$, respectively, and assuming the SLy EoS. The initial frequency of the simulation is set to be $M \omega_{22}^0 = 0.037$, which corresponds to $f = 3.9 \,{\rm kHz}$ and is therefore outside the sensitive bandwidth of current detectors.
They found that extrapolating standard predictions for BH-NS binary models is not reliable in the final phase of the inspiral, close to NS disruption. This affects predictions for the postmerger phase, the associated properties of the ejecta, and remnant properties of the kilonova. 

In this work, we only focused on deviations from the point-particle waveform much before the merger, where the PN approximation on which TaylorF2 is entirely based should be valid. However, the very effect of tidal disruption underscores a departure from the point-particle description of the binary component, making an accurate modeling much more involved. Nevertheless, if properly modeled, any more radical deviation from the TaylorF2 prediction would probably help in ruling out the NS nature of the binary. Therefore, we expect our results to be conservative. On the other hand, it is important to keep in mind that a potential mismatch with the waveform adopted in the data analysis would reduce search sensitivity (see, e.g.~\cite{Bandopadhyay:2022tbi}).

Overall, our results strongly suggest that future observations will be able to distinguish between subsolar PBHs and NSs, but accurate model selection (e.g., by computing Bayes factors between two competitive hypotheses) requires a more precise waveform modeling if one wishes to support strong claims on the nature of putative SSM detections.
Future work should also focus on applying our augmented waveform model in actual searches. By construction we assume zero-noise realizations of the detector sensitivities. The evidence in favor or against the detection of tidal effects may be sensitive to noise realization and may affect borderline individual detections.

Finally, one could also consider stacking~\cite{Pacilio:2021jmq,DelPozzo:2013ala,Lackey:2014fwa} multiple subsolar events to include more information. As the tidal deformability of NSs significantly depends on the mass, one could forecast this procedure only by assuming a mass distribution for the subsolar mergers.  In the case of equally informative and independent observations, we expect the uncertainty on tidal deformability to scale roughly as $1/\sqrt{N_\text{\tiny det}}$. Of course, the advantage of this technique will mostly depend on the putative number $N_\text{\tiny det}$ of detected subsolar events, and the improvement might be relevant only for next-generation detectors.

\vspace{0.5cm}
\textit{Note added.} After this work appeared on the arXiv, we became aware of an upcoming paper, performing a similar but independent analysis~\cite{upcomingpaper}. 
Reference~\cite{upcomingpaper} uses the NRTidalv3 waveform and quantifies the distinguishability of different hypotheses with O4 sensitivity (including mixed BH-NS binaries) by computing the odd ratios.
Despite the two analyses being different in some points, they reach the same results about the measurability of tidal effects in subsolar binaries. This strengthens their conclusion that SSM BHs and NSs can be distinguished in future events with ${\rm SNR}\gtrsim 12$.

\let\oldaddcontentsline\addcontentsline
\renewcommand{\addcontentsline}[3]{}
\begin{acknowledgments}
We thank Ulyana Dupletsa, Francesco Iacovelli, Costantino Pacilio, and Massimo Vaglio for interesting discussions, and Valerio De Luca for comments on the draft. We acknowledge using {\tt GWFast}~\cite{Iacovelli:2022mbg,Iacovelli:2022bbs} for some of the computations performed in this work. 
Some inference analysis has been performed at the Vera cluster supported by MUR and Sapienza University of Rome.
This work is partially supported by the MUR PRIN Grant No. 2020KR4KN2 ``String
Theory as a bridge between Gauge Theories and Quantum Gravity'', by the FARE programme (GW-NEXT, CUP:~B84I20000100001), and by the INFN TEONGRAV initiative.. 
\end{acknowledgments}

\let\addcontentsline\oldaddcontentsline
\appendix
 \section{Parameter estimation methods}\label{app:methods}
\let\oldaddcontentsline\addcontentsline
\renewcommand{\addcontentsline}[3]{}

In this appendix we review the standard parameter estimation methods adopted in the main text.

\subsection{Bayesian inference}

Once a GW signal is detected, statistical approaches are used to extract information about the physical parameters of the source.
This step of GW data analysis is performed within Bayesian inference \cite{bayes1, bayes2, bayes4}. 

In the GW data analysis framework, we are interested in estimating the \textit{posterior distribution} $p(\vec\theta|s)$ of a set of parameters $\vec\theta$, conditioned by the detection of a total signal
\begin{equation}
s(t)=h(t,\vec\theta)+n(t)
\label{output_s}
\end{equation}
where $h(t,\vec\theta)$ is the GW signal, and $n(t)$ is the stationary noise component due to the interferometer(s). The posterior distribution for the hyperparameters $\vec \theta$ can be approximated by
\be\label{pos_dist_F}
p (\vec \theta| s) \propto \pi (\vec \theta) e^{-\frac{1}{2}(h (\vec \theta) - s|h (\vec \theta) - s)}
\ee
in terms of the prior distribution $\pi (\vec \theta)$. The inner product is defined as
\be\label{innprod}
(g|h) = 2\int_{f_\text{\tiny min}}^{f_\text{\tiny max}} \d f \frac{\tilde h (f) \tilde g^* (f) + \tilde h^*(f) \tilde g(f)}{S_n(f)}\,,
\ee
in terms of the Fourier-transformed quantities and the detector noise power spectral density, $S_n(f)$. The frequency band $[f_\text{\tiny min},f_\text{\tiny max}]$ of interest depends on the specific detector. The SNR is given by ${\rm SNR}=\sqrt{(h|h)}$.

The Bayes theorem states then that
\begin{equation}
    p(\vec\theta|s)=\frac{\mathcal{L}(s|\vec\theta)\pi(\vec\theta)}{\mathcal{Z}(s)}
    \label{post}
\end{equation}
where:
\begin{itemize}

    \item $\mathcal{L}(s|\vec\theta)$ is the probability of having a signal $s$ given the (source) parameters $\vec\theta$, and is known as the \textit{likelihood function}; the choice of the likelihood is linked to the noise model that we adopt, for instance, a Gaussian one (see, e.g.,~\cite{Maggiore:2007ulw});

    \item $\pi(\vec\theta)$ indicates the \textit{prior probability distribution} of having the set of parameters $\vec\theta$; it represents our knowledge about $\vec\theta$ before we make the measurement; and
    
    \item $\mathcal{Z}(s)$ is the \textit{evidence}, or marginal likelihood, that is,
\begin{equation}
\mathcal{Z}(s)=\int\mathcal{L}(s|\vec\theta)\pi(\vec\theta){d\theta_1}{d\theta_2}...{d\theta_n}
\end{equation}
where the integral is intended over the full $n$-dimensional parameter space.
\end{itemize}
Computationally, we have carried out Bayesian inference on GW synthetic data with the public software \texttt{BILBY}~\cite{BILBY}. To evaluate the posterior distributions, we used the \texttt{DYNESTY} nested sampling \cite{dynesty}. Moreover, to speed-up the simulations, we implemented the relative binning technique, which well evaluates the likelihood at lower frequencies by expanding it over some fiducial parameters \cite{rb1,rb2}. To have an agnostic analysis as much as possible, we implemented large uniform priors on all the sampled parameters\footnote{The only parameter which we required to have a narrow prior is the chirp mass, in which we chose a width prior of 5$\%$, to improve convergence. This does not affect our results as the chirp mass is the best measured parameter, with much narrower uncertainties than the prior width adopted. Furthermore, we set the spin prior to be uniform in the range $[-0.99,0.99]$.}.

\subsection{Fisher information matrix}
The Fisher information matrix is a basic tool often used to assess the parameter estimation capabilities of GW detectors (see, e.g., Refs.~\cite{PhysRevD.46.5236,PhysRevD.47.2198,Cutler:1994ys,Poisson:1995ef,Berti:2004bd,Ajith:2009fz,Cardoso:2017cfl}, and Refs.~\cite{Vallisneri:2007ev,Rodriguez:2013mla} highlighting the limitations of this approach).

In accordance with the maximum-likelihood estimator principle, we approximate the central values of the hyperparameters at the point $\vec\theta \equiv \vec \theta_\text{\tiny p}$ where the likelihood reaches its peak. In the limit of large SNR, one can perform a Taylor expansion of Eq.~\eqref{pos_dist_F} and get
\be
p (\vec \theta| s) \propto \pi (\vec \theta) e^{-\frac{1}{2}\Gamma_{ab} \Delta \theta^a \Delta \theta^b},
\ee
where $\Delta \vec \theta = \vec \theta_\text{\tiny p} - \vec \theta$ and
\be
\Gamma_{ab} = \lp \frac{\partial h}{\partial \theta^a} \bigg | \frac{\partial h}{\partial \theta^b} \rp_{\vec \theta = \vec \theta_\text{\tiny p}}
\ee
is the Fisher-information matrix. The errors on the hyperparameters are, therefore, given by $\sigma_a = \sqrt{\Sigma^{aa}}$, where $\Sigma^{ab} = \lp \Gamma^{-1}\rp^{ab}$ is the covariance matrix.

Using the waveforms and parameters discussed in the main text, we have computed the Fisher matrix using the public code {\tt GWFast}~\cite{Iacovelli:2022mbg,Iacovelli:2022bbs}.
We include broad Gaussian priors on the parameters 
$\phi_c \in [-\pi, \pi]$ and
$\chi_{1,2} \in [-1,1]$, corresponding to
\be
\label{eq:priors}
\delta \phi_c = \pi, \qquad \delta \chi_{1,2} = 1,
\ee
by adding to the diagonal elements of our Fisher matrix terms of the form $\Gamma^\text{\tiny prior}_{aa} = \Gamma_{aa} +  1/(\delta \theta_a)^2$.

\bibliography{main}
\let\addcontentsline\oldaddcontentsline
\end{document}